\documentclass[12pt,letterpaper]{article}
\usepackage{amsmath,amssymb,array,calc,rotating,epsfig,psfrag,amscd}
\newcommand{\nc}{\newcommand}

 \nc{\ra}{\rightarrow} 
\def\al{\alpha}

\def\eps{\epsilon}
\nc{\veps}{\varepsilon}
\def\gam{\gamma}

\def\om{\omega}
\nc{\vphi}{\varphi}
\def\tha{\theta}
\def\sig{\sigma}

\def\Gam{\Gamma}

\def\Om{\Omega}
\def\Sig{\Sigma}

\nc{\bea}{\begin{eqnarray}}
\nc{\eea}{\end{eqnarray}}
\nc{\be}{\begin{equation}}
\nc{\ee}{\end{equation}}
\nc{\cA}{{\cal A}}
\nc{\cB}{ \cal B}

\nc{\cF}{{\cal F}}
\nc{\cG}{{\cal G}}

\nc{\cL}{{\cal L}}
\nc{\M}{{\cal M}}
\nc{\cM}{{\cal M}}
\def\N{{\cal N}}
\def\cN{{\cal N}}
\def\cO{{\cal O}}
\def\P{{\cal P}}
\def\cP{{\cal P}}
\nc{\cQ}{{\cal Q}}
\nc{\cR}{{\cal R}}

\def\T{{\cal T}}

\def\cV{{\cal V}}
\def\W{{\cal W}}

\nc{\BB}{{\mathbb B}}
\nc{\CC}{{\mathbb C}}
\nc{\DD}{{\mathbb D}}
\nc{\EE}{{\mathbb E}}
\nc{\FF}{{\mathbb F}}
\nc{\GG}{{\mathbb G}}
\nc{\HH}{{\mathbb H}}
\nc{\JJ}{{\mathbb J}}
\nc{\RR}{{\mathbb R}}
\nc{\PP}{{\mathbb P}}
\nc{\QQ}{{\mathbb Q}}
\nc{\ZZ}{{\mathbb Z}}
\nc{\CP}{{\CC\PP}}
\nc{\calone}{{\mathbb 1}}
\nc{\half}{\frac{1}{2}}
\nc{\qrt}{\frac{1}{4}}
\nc{\del}{\partial}

\nc{\delbar}{\bar\partial}
\nc{\Spin}{\operatorname{Spin}}
\nc{\SO}{\operatorname{SO}}

\nc{\Sp}{{\rm Sp}}
\nc{\com}[2]{{ \left[ #1, #2 \right] }}
\nc{\acom}[2]{{ \left\{ #1, #2 \right\} }}
\nc{\rr}{\rightarrow}
\nc{\p}{\partial}
\nc{\LT}{{\LL_\T}}
\nc{\Tr}{{\rm Tr}}
\nc{\tr}{{\rm tr}}
\def\com#1#2{{ \left[ #1, #2 \right] }}
\def\acom#1#2{{ \left\{ #1, #2 \right\} }}
\nc{\tKT}{\widetilde{K3}}
\nc{\ttha}{\tilde{\theta}}
\nc{\tphi}{\tilde{\phi}}
\nc{\tPhi}{\tilde{\Phi}}
\nc{\tpsi}{\tilde{\psi}}
\nc{\tgam}{\tilde{\gam}}
\nc{\tGam}{\tilde{\Gam}}
\nc{\tSig}{\tilde{\Sig}}
\nc{\tc}{\tilde c}
\nc{\te}{\tilde e}
\nc{\tg}{\tilde g}
\nc{\tj}{\tilde j}
\nc{\tp}{\widetilde{p}}
\nc{\tq}{\widetilde{q}}
\nc{\ts}{{\tilde s}}
\nc{\tz}{\tilde z}
\nc{\tD}{{\tilde D}}
\nc{\tE}{{\tilde E}}
\nc{\tG}{{\tilde G}}
\nc{\tH}{{\tilde H}}
\nc{\tM}{{\tilde M}}
\nc{\tN}{{\tilde N}}
\nc{\tP}{{\tilde P}}
\nc{\tQ}{{\tilde Q}}
\nc{\tS}{\tilde{S}}
\nc{\tF}{\tilde{{\cal F}}}
\nc{\tX}{\widetilde{X}}
\nc{\hb}{\hat b}
\nc{\hc}{\hat c}
\nc{\hd}{\hat d}
\nc{\he}{\hat e}
\nc{\hf}{\hat f}
\nc{\hg}{\hat g}
\nc{\hh}{\hat h}
\nc{\hp}{\hat p}
\nc{\hw}{\hat w}
\nc{\hx}{\hat x}
\nc{\hy}{\hat y}
\nc{\hz}{\hat z}
\nc{\hA}{\widehat{A}}
\nc{\hE}{\widehat{E}}
\nc{\hH}{\widehat{H}}
\nc{\hJ}{\widehat{J}}
\nc{\tK}{\widetilde{K}}
\nc{\hM}{\widehat M}
\nc{\hF}{\widehat{\F}}

\nc{\ha}{\widehat \alpha}
\nc{\hphi}{\hat{\phi}}
\nc{\hpsi}{\hat{\psi}}
\nc{\hgam}{\hat{\gam}}
\nc{\hPhi}{\hat{\Phi}}
\nc{\hPsi}{\hat{\Psi}}
\nc{\hGam}{\hat{\Gam}}

\nc{\w}{\wedge}

\nc{\ol}{\overline}
\nc{\abar}{\ol{a}}
\nc{\bbar}{\ol{b}}
\nc{\cbar}{\ol{c}}
\nc{\ebar}{\ol{e}}
\nc{\ibar}{\ol{\imath}}
\nc{\jbar}{\ol{\jmath}}
\nc{\kbar}{\ol{k}}
\nc{\lbar}{\ol{l}}
\nc{\mbar}{\ol{m}}
\nc{\nbar}{\ol{n}}
\nc{\ubar}{\ol{u}}
\nc{\vbar}{\ol{v}}
\nc{\wbar}{\ol{w}}
\nc{\xbar}{\ol{x}}
\nc{\ybar}{\ol{y}}
\nc{\zbar}{\ol{z}}

\nc{\Ebar}{\ol{E}}
\nc{\Jbar}{\ol{J}}
\nc{\Qbar}{\ol{Q}}
\nc{\Wbar}{\ol{W}}
\nc{\Xbar}{{\overline X}}
\nc{\Ybar}{{\overline Y}}
\nc{\Zbar}{{\overline Z}}

\nc{\epsbar}{\ol{\epsilon}}
\nc{\lambar}{\ol{\lambda}}
\nc{\psibar}{\ol{\psi}}
\nc{\Psibar}{\ol{\Psi}}
\nc{\phibar}{\ol{\phi}}
\nc{\Phibar}{\ol{\Phi}}
\nc{\chibar}{\ol{\chi}}
\nc{\ombar}{\ol{\om}}
\nc{\Ombar}{\ol{\Om}}
\nc{\bah}{{\mathbf {\hat{A}}}}
\nc{\bX}{{\mathbf X}}
\nc{\dal}{\dot{\al}}
\nc{\thab}{\bar{\theta}}
\nc{\thal}{\theta^{\al}}
\nc{\thdal}{\bar{\theta}^{\dal}}

\nc{\thsigthm}{\tha \sigma^m \thab}
\nc{\thsigthn}{\tha \sigma^n \thab}

\nc{\Dal}{D_{\al}}
\nc{\Ddal}{\bar{D}_{\dal}}
\nc{\CDal}{{\cal D}_{\al}}
\nc{\CDdal}{\bar{\cal D}_{\dal}}

\nc{\eq}[1]{(\ref{#1})}
\nc{\non}{\nonumber}
\nc{\comment}[1]{{\bf #1}}
\nc{\xs}{\not\!\!X}
\nc{\ps}{\not\!\!P}
\nc{\dif}{{d}}
\nc{\equ}{{\rm eq}}
\def\Im{{\rm Im ~}}
\def\Re{{\rm Re ~}}
\nc{\AdS}{{\rm AdS}}
\nc{\vol}{{\rm vol}}
\nc{\Ainf}{A_{\infty}}
\nc{\End}{{\rm End}}
\nc{\Ext}{{\rm Ext}}
\nc{\Hom}{{\rm Hom}}
\nc{\IIB}{{\rm IIB}}
\nc{\Pic}{{\rm Pic}}
\nc{\bra}[1]{\langle{#1}|}
\nc{\ket}[1]{|{#1}\rangle}
\nc{\braket}[2]{\langle{#1}|{#2}\rangle}
\nc{\sect}[1]{Section~\ref{#1}}
\nc{\fig}[1]{Fig.~\ref{#1}}
\nc{\chap}[1]{Chapter~\ref{#1}}
\nc{\Dslash}{\ensuremath \raisebox{0.025cm}{\slash}\hspace{-0.32cm} D}
\nc{\no}{\!:\!\!}
\nc{\bpm}{\begin{pmatrix}}
\nc{\epm}{\end{pmatrix}}
 \nc{\bitem}{\begin{itemize}}
 \nc{\eitem}{\end{itemize}}

\newcommand{\C}[1]{$(\ref{#1})$}

\def\zb{\bar{z}}
\def\Vol{\operatorname{Vol}}
\def\pz#1{\frac{\p}{\p {z}^{#1}}}
\def\pzb#1{\frac{\p}{\p {\zb}^{#1}}}

\def\cc{\text{c.c.}}

\def\Z{\mathbb{Z}}

\def\ff#1#2{{\textstyle\frac{#1}{#2}}}
\def\SO{\operatorname{SO}}

\nc{\rank}{{\rm rank}} 
\nc{\pr}{{\rm pr}} 

\nc{\tom}{\tilde{\om}} 
\nc{\tOm}{\tilde{\Om}}

\begin{document}

\begin{titlepage}

\begin{center}

%\fbox{DRAFT: \today}

{April 24, 2007} \hfill     EFI-07-11
\vskip 2 cm
{\Large \bf Instantons, Hypermultiplets and the Heterotic String}\\
\vskip 1.25 cm 
{ Nick Halmagyi\footnote{halmagyi@theory.uchicago.edu},  
Ilarion V. Melnikov\footnote{lmel@theory.uchicago.edu} } and Savdeep Sethi\footnote{sethi@theory.uchicago.edu} \\

{ \vskip 0.5cm Enrico Fermi Institute, \\
University of Chicago, \\Chicago, IL 60637, USA\\}

\end{center}

\vskip 0.3 cm

\begin{abstract}
\baselineskip=16pt

Hypermultiplet couplings in type IIA string theory on a Calabi-Yau space can be 
quantum corrected by $D2$-brane instantons wrapping special Lagrangian cycles. On the other hand, 
hypermultiplet couplings in the heterotic string on a K3 surface are corrected by 
world-sheet instantons wrapping curves. In a class of 
examples, we relate these two sets of instanton corrections. We first present an analogue 
of the c-map for the heterotic string via a dual flux 
compactification of M-theory. Using this duality, we propose two ways of capturing quantum corrections to 
hypermultiplets. We then use the orientifold limit of 
certain F-theory compactifications to
relate curves in K3 to special Lagrangians in dual type IIA compactifications. We conclude with some results from perturbative string theory for  hypermultiplet F-terms and a conjecture about the topology of brane instantons.  

\baselineskip=18pt

\end{abstract}

\end{titlepage}

\tableofcontents

%\newpage
%%%%%%%%%%%%%%%%%%%%%%%%%%%%%%%%%%%%
%%%%%%%%%%%%%%%%%%%%%%%%%%%%%%%%%%%%
\section{Introduction}
\label{introduction}

\subsection{BPS couplings in four dimensions}

Compactifications to four dimensions with N=2 supersymmetry are 
quite fascinating because of the extent to which quantum corrections 
to space-time couplings can be understood. In addition to the 
gravity multiplet, the space-time N=2 supergravity theory typically 
contains vector multiplets and hypermultiplets. Each vector multiplet contains
a complex scalar, while each hypermultiplet contains four real scalars organized 
as a quaternion. The moduli space for an N=2
compactification splits locally into a metric product space
\be
{\cal M}_V \times {\cal M}_H.
\ee
These two types of supermultiplet arise 
in quite different ways depending on the string compactification under 
consideration. 
    
The main focus of study has been a class of BPS vector multiplet couplings given in terms of $\cF^{(g)}(z,\bar{z})$
where $(z,\bar{z})$ are vector multiplet moduli; 
see for example~\cite{Antoniadis:1995zn}.  These
couplings appear in the low-energy effective action in the schematic form
\be\label{veccouplings}
\int d^4x \, {\cF}^{(g)}(z, \bar{z}) \, R^2 T^{2g-2} +\ldots
\ee
where $T$ is the graviphoton field strength. The prepotential, $\cF^{(0)}$, determines the metric on the 
vector multiplet moduli space. In a class of examples, these couplings can be determined exactly using
string dualities. The quantum corrections encoded in the $\cF^{(g)}$ can contain interesting information about
instantons like Gromov-Witten invariants of a Calabi-Yau space.    

In this work, we will focus on quantum corrections to the 
hypermultiplet couplings. These couplings are far less well understood. What is known is that the hypermultiplet moduli space  ${\cal M}_H$
is a quaternionic K\"ahler manifold~\cite{Bagger:1983tt}, whose classical form can often be determined. Such a space with quaternion
dimension $n$ has holonomy $Sp(1) \cdot Sp(n)$. In
the limit where gravity is decoupled $G_N \rightarrow 0$, this quaternionic K\"ahler manifold is replaced by a hyperK\"ahler
manifold with holonomy $Sp(n)$.  This local limit is significantly easier to analyze. 

There are several ways of constructing an 
N=2 compactification in string theory. A prime example is type II 
string theory compactified on a Calabi-Yau space $X$. Another 
example is the heterotic or type I string compactified on $K3 \times T^2$. This compactification
requires a choice of gauge bundle. There is also a class of N=2 heterotic compactifications with NS flux~\cite{Dasgupta:1999ss}, and an
assortment of symmetric and asymmetric orbifold compactifications. The cases of interest 
to us in this paper are the heterotic or type I string on $K3 \times 
T^2$ and the type IIA string on a Calabi-Yau space $X$. These two compactifications can be
dual if $X$ admits a $K3$ fibration.

In many regards, the natural setting in string theory for a study of hypermultiplet 
couplings is the heterotic string. As illustrated in 
table~\ref{variouscouplings}, the hypermultiplet couplings computed in the 
heterotic string are not renormalized by either string loop or string 
non-perturbative effects. This follows because the heterotic dilaton is part of a 
vector multiplet. 

The hypermultiplet couplings are therefore determined by $(0,4)$ superconformal field
theory. When computed in terms of a non-linear sigma model, the hypermultiplet
couplings can receive both perturbative $\alpha'$ corrections and corrections from heterotic
world-sheet instantons. At large volume, these world-sheet instanton effects are  
counted by a generalization of Gromov-Witten theory which takes into account 
the bundle structure. These instanton corrections are therefore naturally studied
in complex geometry.

\begin{table}[bth!]
\begin{center}
\begin{tabular}{|c|c|c|}
\hline
& Vector & Hyper \\
\hline 
IIA on $X$ & $\vol(X)$ & $g_s$\\
\hline
Heterotic on $K3\times T^2$ & $g_s$ & $\vol(K3)$\\
\hline
\end{tabular}
\end{center}
\caption{The type of supermultiplet containing the string coupling and 
the volume which controls $\alpha'$ corrections  in type IIA and 
heterotic string theory.} \label{variouscouplings}
\end{table}

On the other hand, in type II string theory on $X$ hypermultiplet couplings can receive both perturbative and
non-perturbative corrections in the string coupling. In type IIA string theory, there are no $\alpha'$ corrections
but there are quantum corrections from Euclidean D2-branes wrapping 
special Lagrangian submanifolds of $X$ (sL 3-folds or sLags)~\cite{Becker:1995kb}. There are also quantum corrections from 
Euclidean NS5-branes wrapping $X$ and bound states of both branes.  Little 
is actually understood about the structure of 
sLags, or how to appropriately count them when they appear in families; see~\cite{joyce-2003}\ for some recent discussion. 
The main difficulty with studying sLags is that they involve real rather than complex geometry. 

In string theory, BPS hypermultiplet couplings provide a natural counting function for these various
instanton configurations. The idea of counting will, in general, involve more than just the semi-classical BPS
configurations. This is necessarily the case because couplings in the space-time effective action vary smoothly with moduli, even when crossing 
curves of marginal stability. BPS particles, however, can decay across such curves. Therefore, space-time couplings receive contributions
from Euclidean space BPS instanton configurations together with some non-BPS contributions needed for smoothness. Usually it is possible to 
unentangle the two types of contribution. This phenomenon is already well studied in
supersymmetric field theory.

So, for example, the hypermultiplet moduli space metric ${\cal G}$ which determines the space-time kinetic terms 
for the hypermultiplets $q$,
\be 
\int d^4 x \, {\cal G}_{ij}(q) D_\mu q^i D^\mu q^j + \ldots,
\ee   
counts sLag and NS5-brane instanton corrections when computed in type IIA on $X$ at weak coupling. When computed in the heterotic string on $K3\times T^2$ at large volume, this metric counts world-sheet instantons.  However, actually computing these quantum corrections has proven to be challenging in any compactification. 

Now at weak coupling in type IIA or large volume for the heterotic string, there is a distinguished hypermultiplet, $q_Z$, known as the universal hypermultiplet. It contains either the string dilaton or the volume modulus in type II or heterotic string theory, respectively. Using $q_Z$, one can construct the hypermultiplet analogue $\tF^{(g)}(q)$ of the higher derivative vector couplings~\C{veccouplings}~\cite{Antoniadis:1993ze}. These couplings will be described later in section~\ref{ss:Hypermultiplets}. It is important to stress that at generic points in ${\cal M}_H$, there is no distinguished hypermultiplet. 

What has been studied so far is the one string loop correction to the metric on $ {\cal M}_H$ for type II on $X$ which is proportional to $\chi(X)$~\cite{Antoniadis:1997eg, Antoniadis:2003sw}. In addition, quantum corrections in the type II string have been studied in~\cite{Davidse:2003ww, Anguelova:2004sj, Davidse:2004gg,  deVroome:2006xu, Robles-Llana:2006is}. A particularly well examined case is when $ {\cal M}_H$ is low-dimensional. In this situation, the quaternionic K\"ahler condition becomes much more powerful~\cite{Alexandrov:2006hx, Robles-Llana:2006ez, deWit:2007qz}.  

\subsection{BPS couplings in three dimensions}

Now there are good reasons to expect the structure of hypermultiplet couplings to be as beautiful as that of vector multiplets. On a further circle
compactification to three dimensions, hypermultiplets and vector multiplets become equivalent. The way this works goes as follows: a $4$-dimensional abelian vector $A_\mu$ compactified on $S^1$ gives rise to one compact scalar from the Wilson line on $S^1$,
\be \int_{S^1} A_\mu dx^\mu, \ee
and a second compact scalar arises from dualizing the three-dimensional photon. Similarly, the circle reduction of the N=2 gravity multiplet  (graviton and graviphoton) gives rise to one extra hypermultiplet. Under this classical duality, the special K\"ahler moduli space ${\cal M}_V$ of complex dimension $n_V$ is replaced by a quaternionic K\"ahler moduli space $ {\widetilde{\cal M}}_H$ of quaternionic dimension $n_V+1$.  Decoupling the hypermultiplet from the gravity sector yields a hyperK\"ahler rather than quaternionic K\"ahler space. The moduli space $ {\widetilde{\cal M}}_H$ obtained this way admits a torus action for each resulting  hypermultiplet. 

On compactification to three dimensions, $ {\cal M}_H$ is unchanged. One can now imagine exchanging $ {\widetilde{\cal M}}_H$ with $ {\cal M}_H$. 
This exchange is realized in the type II string by T-duality. Type IIA on $X \times S^1$ is equivalent to type IIB on ${X\times {\hat S^1}}$ where $S^1$ and $\hat{S}^1$ are related by T-duality. The resulting classical map between moduli spaces is known as the c-map~\cite{Cecotti:1988qn}. The explicit mapping of classical supergravity actions has been worked out in~\cite{Ferrara:1989ik}. Some aspects of the geometry of this map together with an attempt to include string loop effects appear in recent work~\cite{Rocek:2005ij}. It is important to stress that the T-duality we are describing is, at best, a perturbative symmetry of the type II string on  $X \times S^1$.   

Unfortunately, as soon as we compactify to three dimensions, it is no longer possible to easily determine the complete set of quantum corrections to the vector multiplet couplings. There are new instanton effects in three dimensions which correspond to monopole gauge-field configurations. The effects of these monopole instantons are difficult to determine exactly even when gravity is decoupled. These instantons break the torus action on  $ {\widetilde{\cal M}}_H$ to a discrete subgroup. For example, in type IIB on $X\times S^1$ a subset of these monopole instantons are constructed from Euclidean D3-branes wrapping sLags in $X$ and wrapping the $S^1$ factor. 

In addition, there is a particularly problematic instanton which arises as follows: in both type II on $X\times S^1$ and heterotic on $K3\times T^2 \times S^1$, there is a new gauge field arising from the metric reduced on $S^1$. Monopoles for this gauge-field are constructed as gravitational instantons in both string theories. The action for this instanton is non-perturbative in the three-dimensional string coupling in both compactifications. If there were a method to capture this instanton and exactly determine the vector multiplet moduli space in three dimensions then one could hope to determine the exact quantum corrected metric on ${\cal M}_H$. This might be possible using modular properties of the vector multiplet couplings~\cite{Huang:2006hq, Grimm:2007tm}\ suitably extended to three dimensions together with the determinable subset of quantum corrections that we will describe.     

\subsection{Plan and summary}

We begin in section~\ref{reviewslag}\ by reviewing some facts about sLag geometry. We pay particular attention to sLags in tori, $K3$ surfaces and in Borcea-Voisin $3$-folds. 

In section~\ref{hetduality},  we first describe a duality between two different compactifications of the heterotic string on $K3\times T^3$. This duality follows from a study of M-theory on $K3\times \widetilde{K3}$ with flux. The resulting map provides an analogue of the c-map for the heterotic string which does not rely on a perturbative symmetry like T-duality.
Along the way, we will encounter quantum phenomena like emergence of a new dimension in the heterotic string from wrapped NS5-branes. 

Using this map, we propose two concrete ways of controlling quantum corrections to the heterotic string hypermultiplet moduli space. The first is via a perturbative computation of the heterotic vector multiplet couplings in three dimensions. This captures quantum corrections to hypermultiplet couplings in the large volume limit of the $K3$ surface. In a dual type IIA compactification, this computation maps to weak string coupling in which we expect to distinguish loop effects from sLag instantons and NS5-branes.  

The second approach is via the $\cF^{(g)}$ vector couplings computed in type IIA string theory on a $K3$-fibered space $X$. The limit of interest is where the volume, 
\be
{\rm vol} (\P^1) \rr 0,
\ee
where $\P^1$ is the base of the $K3$ fibration. This is the opposite of the limit usually studied. In this limit, we capture a different set of quantum corrections to the hypermultiplets in the dual heterotic and type IIA compactifications.   

For a class of models, there is a fairly explicit map at the orientifold locus from the heterotic theory to type IIA. In section~\ref{stringcurves}, we apply this map to instanton configurations and find that world-sheet instantons in $K3$ can be mapped to brane instantons in $X$. Along the way, we encounter a localization phenomenon where the contribution of many sLags cancel out of the hypermultiplet couplings. Instead, the sum over sLag instantons can be replaced by a sum over sLags with special structure. We also find that an obstruction to wrapping cycles in a homology class with ``no vector structure'' in type I or $Spin(32)/\ZZ_2$ heterotic is realized geometrically in the type IIA dual by the absence of any corresponding sLag.   

In section~\ref{fterms}, we derive some selection rules for computing the $\tF^{(g)}(q)$ couplings in heterotic and type I perturbative string theory. We find that $\tF^{(g)}(q)$ is perturbatively $g$-loop exact in both the heterotic and type I string. In addition, the $\tF^{(g)}(q)$ are generated at a fixed order in $\alpha'$ perturbation theory in the heterotic string.

We conclude in section~\ref{topology} with a discussion of the topological complexity of a sLag $L$. Based on the structure of fermion zero modes, we arrive at a conjecture that any sLag $L$ which is dual to a world-sheet instanton of genus $g$ satisfies the relation
\be
b_1(L) \geq g, 
\ee
where $b_1$ is the first Betti number of $L$. We also show that for the case of the standard embedding, the $\tF^{(g)}(q)$ couplings are not renormalized by world-sheet instantons in the heterotic string.

\section{Special Lagrangian Submanifolds}

Let us begin by reviewing the definition of special Lagrangian submanifolds or sLags. We will need particular 
cases in our analysis: namely sLags in $2$-folds and $3$-folds but we will begin with the general case. 

\label{reviewslag}
\subsection{Basic definitions and properties}

Let $X$ be a Calabi-Yau $n$-fold with a K\"ahler form $\omega$, a Ricci-flat metric $g$, and a holomorphic $(n,0)$ form $\Omega$.
A submanifold $L \subset X$  is said to be {\em Lagrangian} if
$\omega|_L = 0$.\footnote{This short-hand notation should be read as
follows: splitting the tangent bundle $TX|_L$ as $TL\oplus NL$, we demand
that for any $p \in L$ and any $v,w\in TL_p \subset TX_p$ we have $\omega(v,w) = 0$.} 
Our interest is mainly in {\em special Lagrangians} (sLags) $L$ which satisfy the extra condition
that ${\rm Im}\left( e^{i\theta} \Omega \right) =0$ for some constant angle $\theta$.\footnote{The careful reader will note that for a single sLag $L$ the phase $\theta$ is entirely irrelevant because it can be absorbed into $\Omega$.  However, it does play an important role for a discussion of A-brane stability \cite{Aspinwall:2004jr}.} There  are several features of sLag geometry that will be important in what 
follows:
\begin{enumerate}
 \item  Since $\omega$ is a non-degenerate form on $X$, $\dim_{\RR} L =n$, and $\omega$ yields an isomorphism from $T L$ to $N^{\ast} L$.
 \item  $L$ is calibrated by ${\rm Re}\left(e^{i\theta} \Omega \right)$, and, hence, is  absolutely volume minimizing with respect to the metric $g$.  
 \item The local deformation theory of a sLag is  extremely simple:  it is a theorem of McLean\cite{McLean:1998} that 
          the deformations of a sLag $L$ are in one-to-one correspondence with the harmonic $1$-forms on $L$. Locally the deformation moduli space of $L$ is a smooth manifold  of dimension $b_1(L)$.
\item  The K\"ahler form calibrates holomorphic submanifolds of $X$, so $\omega|_L=0$ also implies that $T_p L$ cannot contain a holomorphic 
           plane.
\end{enumerate}
These properties are discussed in greater length in a very lucid review by Joyce~\cite{Joyce}, but hopefully
this brief reminder will be sufficient.  We will now give several examples of sLag geometry that 
will be relevant for our work.

\subsection{Special Lagrangian cycles in $T^2$}
Since $T^2$ is a particularly simple space, one might expect that sLags in $T^2$ admit a particularly simple description.  This is
indeed the case. We can describe a flat $T^2$ as a quotient of $\CC$ by the identifications: $z \sim z+1$ and $z \sim z + \tau$ for some $\tau$ in 
the upper half-plane. Writing $z= x+ \tau y$, with $x,y$ normalized with period one, the natural K\"ahler and holomorphic one-form are
\begin{eqnarray*}
\omega  &=& \tau_2 \, dy \wedge dx,\\
\Omega &=& dz = dx + \tau dy.
\end{eqnarray*}
Parameterizing a one-dimensional submanifold $L$ by $(x(t),y(t))$, it is clear that the Lagrangian condition is vacuous, while the
special condition with angle $\theta$ requires $L$ to be a straight line in the $T^2$ with 
\be 
x = - (\cot(\theta) \tau_2 + \tau_1) \, y. 
\ee
Note that in order for this to be a well-defined cycle $\cot(\theta) \tau_2 + \tau_1$ must be rational, giving a constraint on possible
values of $\theta$.  It is easy to generalize this construction to higher dimensional tori $T^{2n}$, but not all sLags will be obtained 
in this fashion.  For example, one can find sLags in $T^6$ that are not homeomorphic to a $T^3$\cite{DanLee}.

\subsection{sLags in $K3$} \label{sec:sLagK3}
Since we do not know any smooth Ricci-flat $K3$ metric, the sLags in a $K3$ surface $S$ are harder to describe than the sLags on tori.  However, 
the hyper-K\"ahler structure of $S$ does allow us to relate sLags in a given complex structure to holomorphic curves in different complex structure 
\cite{Joyce,HarveyLawson:1982}.  This is quite important, as it translates a problem in real geometry to a (hopefully) more tractable problem in
complex geometry.  Of course, finding holomorphic curves is by no means an easy task!

To see this, recall that $S$ admits three anti-commuting complex structures $I,J,K$ with $IJ=K$, compatible with the Ricci-flat metric.
Fixing a given complex structure, say $I$, we can write $\omega$ and $\Omega$ in terms of self-dual two-forms on $S$:  $\omega = s_1$,
$\Omega = s_2 + i s_3$, where the $s_i \in H^{2}(S,\RR)$ satisfy 
\be
s_i \wedge s_j = 2 \Vol_g (S) \, \delta_{ij}. 
\ee
The $s_i$ transform in the adjoint 
representation under $SU(2)$ rotations of the complex structures $I,J,K.$  Suppose $L$ is a sLag with phase $\theta$ in complex 
structure $I$.  This implies that 
\begin{eqnarray*}
\omega|_L & = & 0\\
\sin\theta \, {\rm Re}(\Omega)|_L + \cos \theta \, {\rm Im}(\Omega)|_L & = & 0 \\
\cos\theta \, {\rm Re} (\Omega) |_L - \sin\theta \, {\rm Im}(\Omega)|_L & = & \Vol_g (L).
\end{eqnarray*}
Now consider the following $SU(2)$ rotation of complex structures:
\begin{equation*}
\left( \begin{array}{c} I' \\ J' \\ K' \end{array} \right) = 
\left( \begin{array}{ccc} 0 & \cos\theta & -\sin\theta \\ 0 & \sin\theta & \cos\theta \\ 1 & 0 & 0 \end{array} \right) 
\left( \begin{array}{c} I \\ J \\ K \end{array} \right).
\end{equation*}
Under this rotation we find $\omega'|_L = \Vol_g (L)$, while $\Omega'|_L =0$.  This is, of course, just the condition for 
$L$ to be holomorphic in complex structure $I'$. 

Having reduced the study of sLags in $S$ to the problem of finding holomorphic curves in $S$, let us comment on the
structure of these ``simpler'' algebraic objects.  As a start, one might ask which cycles in $H_2(S,\ZZ)$ are representable by
holomorphic curves.  We will now argue that given $C\in H_2(S,\ZZ)$ either $C$ or $-C$ has a holomorphic representative,
and hence, either $C$ or $-C$ has a sLag representative for some choice of the $SU(2)$ rotation of complex structures.

The proof is well-known, and we merely give it for completeness.  Let $C$ be a divisor in $H_2(S,\ZZ)$ with $C\cdot C = 2g-2$, and denote by $L_C$ the line bundle
corresponding to $C$.  The Riemann-Roch formula for surfaces may be used to compute the holomorphic Euler 
characteristic of $L_C$:
\begin{equation}
\chi(L_C) = h^0(L_C) - h^1(L_C) + h^2(L_C) = 1+g.
\end{equation}
Here $h^i(L_C)$ are the dimensions of $H^i(X,\cO (L_C))$, and in particular, $h^0(L_C)$ counts the number of holomorphic sections of $L_C$.  
Furthermore, Serre duality on $S$ implies $h^2(L_C) = h^0(L_{-C})$, and we obtain
\begin{equation}
h^0(L_C) - h^1(L_C) + h^0(L_{-C}) = 1+g.
\end{equation}
Thus, either $L_C$ or $L_{-C}$ is a positive line-bundle.  Suppose $L_C$ is positive.  Then the Kodaira vanishing theorem asserts 
that $H^q(X, \Omega^n(L_C)) =0$ for $q+n>2$.  But, recall that $\Omega^2(L_C) \simeq L_C$ on $S$, since $S$ has trivial canonical 
class. It follows that $h^1(L_C) = h^2(L_C) = 0$.  Alternatively, if $L_{-C}$ is positive it follows that $h^1(L_{-C}) = h^2(L_{-C})=0$.  

Thus, corresponding to either $C$ or $-C$ there exists a line bundle with $1+g$ holomorphic sections.  The corresponding holomorphic curve 
may be taken to be the zero of one of these sections.  For $g >0$ we have 
rather an embarrassment of riches:  there is, modulo obstructions, a whole family of curves.  In fact, for $g >1$ it is even possible for the 
holomorphic curve to have a number of disconnected components!\footnote{We thank Paul Aspinwall for clarifying this point.}

\subsection{sLags in 3-folds}
Special Lagrangian submanifolds of Calabi-Yau 3-folds have a much richer structure.  From the discussion of
sLags above, it is clear that we can construct many sLags in three-folds with degenerate holonomy.  For example,
we can easily write down many sLags in $T^6$ that are homeomorphic to $T^3$, and similarly, we can identify
many sLags of the form $C \times S^1$ in $S \times E$, where $S$ is a $K3$ surface and $E$ is an elliptic curve. 
In this case, $C$ is a holomorphic cuve in $S$ and $S^1$ is a cycle in $E$.  

$T^6$ and $S\times E$ are examples of trivial elliptically-fibered
Calabi-Yau 3-folds. More non-trivial (singular) examples of 3-folds can be obtained  by orbifolding these
product spaces.  Under favorable conditions,  
the trivially constructed sLags in the product spaces will descend to 
sLags in the quotient space.   As we will show, heterotic/IIA duality implies that these ``trivial'' sLags play a distinguished
role in instanton corrections.  However, it is important to realize that these are very special, and most of the sLags in the 
product space are not of this form.  We will now illustrate this in a concrete example of $T^3$ sLags in $T^6$.

\subsubsection{Some sLags in $T^6$}
For simplicity, we work with a $T^6$ that is the product of three square tori with coordinates $z^i = x^i+i y^i$ and K\"ahler
and holomorphic forms
\begin{equation*}
\omega = \frac{i}{2} \sum_{i=1}^3 d z^i \wedge d \bar{z}^i, ~~~~ \Omega = dz^1 \wedge dz^2 \wedge dz^3.
\end{equation*}
We consider a set of $3$-cycles $L$ constructed as graphs $T^3 \to T^3$,
\begin{eqnarray}
\label{eq:L}
y^1  & = & \alpha x^1 + \beta x^2 + \gamma x^3, \nonumber\\
y^2  & = & \beta x^1 + \delta x^2 + \eps x^3, \nonumber\\
y^3  & = & \gamma x^1 + \eps x^2 + \rho x^3,
\end{eqnarray}
where $\alpha,\ldots,\rho$ are integer parameters.  The tangent space to $L$ is spanned by
\begin{eqnarray}
\label{eq:TL}
V_1  & = & (1+i\alpha) \pz1 + i \beta \pz2 + i \gamma \pz3 + \cc, \nonumber\\
V_2  & = & i \beta \pz1 + (1+i\delta) \pz2 + i \eps \pz3 + \cc, \nonumber\\
V_3  & = & i\gamma \pz1+ i\eps  \pz2 + (1+i\rho) \pz3 + \cc.
\end{eqnarray}
Note that the $V_A$ are linearly independent for all values of the parameters.

$L$ is automatically Lagrangian since $\omega(V_A,V_B) = 0$.  In order for $L$ to be
sLag (with $\theta=0$), we also need $\Im \Omega(V_1,V_2,V_3) =0$.  Of course,
\begin{equation}
\Omega(V_1,V_2,V_3) = \left| \begin{array}{ccc}   (1+i\alpha) &    i\beta       &    i\gamma \\
                                                                                               i\beta     & (1+i\delta) & i\eps            \\
                                                                                             i\gamma & i\eps             & (1+i\rho)  
                                                      \end{array} \right|,
\end{equation}
or
\begin{eqnarray}
\Re \Omega(V_1,V_2,V_3) & = & 1 + \beta^2+\gamma^2+\eps^2 -\alpha\delta -\alpha\rho  -\delta\rho, \\
\Im \Omega(V_1,V_2,V_3) & = & \alpha+\delta+\rho +\alpha \eps^2+\delta\gamma^2+\rho\beta^2 -\alpha\delta\rho -2 \beta \eps\gamma. \nonumber
\end{eqnarray}
There are many integer solutions to $\Im \Omega = 0$.  As a simple example, we may set 
$\eps=\delta=0$, and then determine $\alpha$ in terms of $\beta,\gamma,\rho$:
$\alpha = -\rho -\rho\beta^2$.

Consider the ``elliptic fiber'' $E$ with tangent directions spanned by $\pz1$ and $\pzb1$.  We will now
show that $TE$ and $TL$ need not have a non-trivial intersection.  Such an intersection would mean
that there exist constants $C^A$, not all zero, such that $\sum_A C^A V_A^i = 0$ for $i =2,3$.  More
explicitly, we need
\begin{eqnarray}
\label{eq:tosolve}
C^1 i\beta + C^2(1+i\delta) + C^3 i\eps & = & 0, \nonumber\\
C^1 i\gamma + C^2 i\eps + C^3 (1+i\rho) & = & 0.
\end{eqnarray}
These equations imply that $C^2=C^3=0$, so that unless $\beta =\gamma=0$, $T L$ cannot intersect
$TE$.  Thus, the sLags with non-zero $\beta$ or $\gamma$ will not respect the decomposition $T^6 \simeq T^4 \times E$.

There are, of course, sLags that do wrap a cycle in $E$:  setting $\beta=\gamma=0$, we find
\begin{equation}
\Im \Omega(V_1,V_2,V_3)  = \alpha+\delta+\rho +\alpha \eps^2 -\alpha\delta\rho =0.
\end{equation}
Any solution of this relation gives such a sLag.  

\subsubsection{Product sLags in Borcea-Voisin three-folds} \label{sLagBV}

Borcea-Voisin (BV) manifolds \cite{Borcea:1994, Voisin:1992} are elegant constructions of Calabi-Yau $n$-folds from two lower dimensional Calabi-Yau manifolds. In general one starts with the pair of Calabi-Yau manifolds $(M_1,M_2)$ of dimension $(m_1,m_2)$ (of course $n=m_1+m_2$) with holomorphic involutions $(\sig_1, \sig_2)$. These involutions must reverse the sign of the respective holomorphic top-forms $(\Omega_1,\Omega_2)$ and were first considered in certain cases by Nikulin \cite{Nikulin}. The Borcea-Voisin manifold is
\be
X=\left(M_1\times M_2\right) /\left(\sig_1,\sig_2\right)
\ee
and is clearly Calabi-Yau since its holomorphic top-form 
\be
\Omega=\Om_1\w \Om_2
\ee
 is invariant under the quotient. This construction  yields a BV manifold at a singular point in its K\"ahler moduli space, and a smooth
 three-fold may be obtained by blowing up the singularities by moving into the interior of the K\"ahler cone.
 
It is possible to construct a sLag in $X$ from a pair of sLags in $M_1,M_2$. Suppose that $(L_1,L_2)$ are sLags in $(M_1,M_2)$ with phases 
$(\phi_1,\phi_2)$ so that
\bea
&& \Im \left( e^{i\phi_1} \Om_1\right)|_{L_1}=0 \\
&& \Im \left( e^{i\phi_2} \Om_2\right)|_{L_2}=0. 
\eea
Now since
\bea
\Im (e^{i(\phi_1+\phi_2)} \Om) &=&\Im \left( e^{i\phi_1} \Om_1\right) \w \Re \left( e^{i\phi_2} \Om_2 \right) \non \\
&+&\Re \left( e^{i\phi_1} \Om_1\right) \w \Im \left( e^{i\phi_2} \Om_2 \right)
\eea
clearly $L=L_1\times L_2\subset X$ is a sLag of phase $\phi=\phi_1+\phi_2$. If $(L_1,L_2)$ are smooth and do not intersect the
fixed points of  $(\sig_1,\sig_2)$, then $L$ will be smooth.

The case of interest to us is $(M_1=K3,M_2=T^2)$. The $\sig_1$ quotient actions were classified by Nikulin~\cite{Nikulin} and 
$\sig_2$ is the involution $z\ra -z$.

As explained in section \ref{sec:sLagK3}, for every $C \in H_2(S,\ZZ)$ either $C$ or $-C$ will have a sLag representative for some
$SU(2)$ rotation of the complex structure.  Clearly any two classes $C$ and $C'$ that can be made holomorphic by the same $SU(2)$
rotation will have sLag representatives with the same phase.  A particular case of this observation was used to demonstrate that the SYZ fibrations~\cite{Strominger:1996it} of $K3$ and $T^2$ give rise to the 
SYZ fibration of $X$~\cite{gross-1996}.

\section{A Heterotic/Heterotic Duality}
\label{hetduality}

We would now like to describe a duality between heterotic string compactifications that will lead to a computational framework for determining quantum corrections to hypermultiplet couplings via calculations involving vector multiplets. We begin by recalling the basic structure of the germane compactifications.

\subsection{The basics of heterotic/type I  on $K3 \times T^2$}
\label{heteroticreview}

To describe a heterotic string compactification, we need to specify a target space metric and a gauge bundle $E$. 
We will take our target space to be $K3 \times T^2$ or  $K3 \times T^2 \times S^1$ together with an $E_8\times E_8$ or $Spin(32)/\Z_2$ gauge bundle. As in the prior section, we will denote our $K3$ surface by $S$.  In the large volume limit, the gauge bundle must have self-dual field strength with instanton number $24$. Therefore the gauge bundle $E$ must be holomorphic with a field strength that is a $(1,1)$ integral class on $S$ for all choices of complex structure. 

In the  $E_8\times E_8$
case, the distribution of instantons between the two $E_8$ factors is specified by $(n_1, n_2)$
where 
\be\label{hetbianchi}
n_1+n_2=24 - n_5.
\ee 
The number $n_5$ specifies the number of NS5-branes at points on the
$K3$ surface. If $n_5=0$, the compactification is perturbative. For the most part, we will restrict to this
case.  If the compactification does not originate from a compactification in six dimensions then some of the instantons might be embedded in gauge-fields arising from the torus factor, or there
might be NS flux.  For the moment, we will restrict to compactifications that 
originate from six dimensions.

The $n_H$ hypermultiplets of this compactification correspond to 
moduli of the world-sheet conformal field theory. At large volume, these moduli describe unobstructed
geometric moduli of $S$ together with moduli that describe deformations of the holomorphic gauge bundle $E$. In general, many of the geometric moduli of $S$ will be obstructed by the holomorphicity condition on $E$.

The heterotic string non-linear sigma model is believed to flow 
to a $(0,4)$ superconformal field theory.  The sigma model metric is known to receive 
perturbative  $\alpha'$ corrections for all choices of bundle except $E= TS$~\cite{Howe:1992tg}. In this  
exceptional case, the world-sheet theory enjoys $(4,4)$ supersymmetry and the sigma model metric is uncorrected.

The universal hypermultiplet $q_Z$ arises from the volume modulus of $S$ 
together with three scalars obtained from the $B$-field reduced on the three  self-dual forms of $S$. 
Together these four scalars form a hypermuliplet 
which is distinguished at large volume and present for any compactification of this kind. Note that the $T^2$ factor plays no role in the local structure of the hypermultiplet moduli space. Rather there are $n_V+3$ 
vector multiplets where $n_V$ is the number of vector multiplets present in six dimensions.  For perturbative compactifications, there is a constraint on the difference
\be
n_H - n_V = 244-3. 
\ee     
Since the four-dimensional string coupling is part of a vector multiplet, ${\cal M}_H$ is exact at tree-level and given by the
Zamolodchikov  metric of the $(0,4)$ conformal field theory. From the perspective of the non-linear sigma model, this metric receives both perturbative and non-perturbative corrections in $\alpha'$. 

The non-perturbative corrections come from world-sheet instantons which correspond to curves $C \in H_2(S, \Z)$ which are holomorphic in some complex structure. Not all curves contribute to a given space-time coupling. For genus $0$ curves, the bundle $E$ restricted to $C$ must trivialize to avoid extra left-moving fermion zero modes killing the contribution of the curve~\cite{Distler:1986wm}. 

For the $Spin(32)/\Z_2$ case, the topology of the gauge 
bundle is further characterized by a generalized Stiefel-Whitney class ${\widetilde w}_2$ that describes
compactifications with  ``no vector structure''~\cite{Berkooz:1996iz}. There are three distinct cases on $S$:
\be \label{nvscases}
{\widetilde w}_2 \cdot {\widetilde w}_2=0 \,\, {\rm mod} \, \, 4, \qquad {\widetilde w}_2 \cdot {\widetilde w}_2=2 \, \, {\rm mod} \, \, 4, \qquad {\widetilde w}_2=0.
\ee
For a curve to contribute to the renormalization of the metric
on ${\cal M}_H$, the restriction of ${\widetilde w}_2$ to $C$ must also be trivial~\cite{Witten:1999eg}. 

Ideally, we would like to sum up all such curves and $\alpha'$ corrections and directly determine
the exact quantum corrected ${\cal M}_H$. This might be possible by extending recent developments in summing $(0,2)$ heterotic instantons~\cite{Adams:2003zy, Katz:2004nn, Sharpe:2005fd, Sharpe:2006qd}. However, there is a basic difficulty
because world-sheet techniques are naturally adapted for holomorphic quantities and ${\cal M}_H$ is not naturally expressed in terms of a holomorphic prepotential except in local limits where gravity decouples. Even these local cases are interesting~\cite{Sen:1997js, Witten:1999fq}\ but we will not pursue them here.   

Now the type I and $Spin(32)/\Z_2$ heterotic strings are S-dual to each other in ten dimensions. On reduction to four dimensions, the four-dimensional string couplings and volumes are related as follows
\be\label{4dcouplingrelation}
(g_s^H)^2 = \frac{g_s^I}{\sqrt{V^I}}, \qquad V^H  = \frac{1}{ (g_s^I)^3 \sqrt{V^I}}.
\ee
Here $V$ refers to the six-dimensional compactification volume. The superscripts $H,I$ specify heterotic or type I parameters respectively, and there are similar relations with $H,I$ reversed~\cite{Antoniadis:1997nz}. Note that this
is a weak-weak duality in regimes of parameter space in four dimensions.

Unlike the heterotic string, the type I string is described by a $(4,4)$ superconformal field theory. Effects
from the gauge bundle or open strings arise at higher loops in the string coupling. For the type I string, the volume of $S$
is part of a vector multiplet so there are no $\alpha'$ corrections to ${\cal M}_H$. However, there are both string loop corrections and D-instanton corrections to ${\cal M}_H$. Under this duality map, world-sheet heterotic string instantons map to Euclidean D1-branes of the type I string while perturbative $\alpha'$ corrections map to string loop corrections.

\subsection{Heterotic/M-theory duality}

We would now like to present an analogue of the c-map for the heterotic string. We will then examine the implications of this map for the dual type IIA compactifications. There are a number of interesting related observations in past work~\cite{Sethi:1996es, Perevalov:1997ht, Perevalov:1998eg, Aspinwall:2005qw, Berglund:2005dm}.  

Our starting point is M-theory on $K3\times K3'$. Let us denote the first $K3$ surface by $S$ and the second by $S'$. There is an M2-brane tadpole in this background that must be canceled~\cite{Becker:1996gj, Sethi:1996es}. The cancelation can be accomplished by a combination of  $n_{M_2}$ inserted M2-branes and $4$-form flux $G_4$ satisfying the tadpole constraint~\cite{Dasgupta:1999ss}
\be\label{Mtadpole}
\frac{1}{ 2} \int{ \frac{G_4 }{ 2\pi} \wedge \frac{G_4 }{ 2\pi}} + n_{M_2} = 24. 
\ee
A basic requirement is that 
\be \frac{G_4}{ 2\pi} \in H^{2,2}(S \times S', \Z) \ee 
and that $G_4$ be primitive. We can construct suitable fluxes as follows, 
\be
\frac{G_4 }{ 2\pi} = \omega \wedge \omega', 
\ee 
where $\omega \in H^2(S, \Z)$ while $\omega' \in H^2(S', \Z)$. The amount of supersymmetry preserved by this compactification depends on the choice of $G_4$. If $G_4$ is primitive with
respect to each of the $\PP^1 \times \PP^1$ choices of complex structure of $S \times S'$ then the full N=4 supersymmetry is preserved; examples of this kind can be found in~\cite{Dasgupta:1999ss}. Otherwise only N=2 supersymmetry is preserved. 

We will restrict to cases where the full N=4 is preserved. In these cases, there are two sets of hypermultiplets from $S$ and $S'$ charged with respect to distinct $SU(2)_R$ symmetries. The moduli space is locally a product of quaternionic K\"ahler manifolds $\M_S \times \M_{S'}$.

\subsubsection{The basic duality in seven dimensions}
We now make use of the $7$-dimensional duality~\cite{Witten:1995ex}\ between 
\begin{equation}
\text{Het on}~T^2\times S^1_R \sim \text{M-theory on}~S',
\end{equation}
where $S'$ is an elliptically fibered K3 surface with  fiber $E'$ and section $B'$ and
total volume $V'$. We will also use $E'$ and $B'$ to denote the volume of the fiber and section, respectively.
The choice of elliptic fibration corresponds directly to picking a circle in $T^3$
since both choices distinguish a  $\Gamma^{1,1}$ factor\footnote{We will always use $\Gam^{1,1}$ to denote the unique even self-dual lattice of signature $(1,1)$. See section~\ref{stringcurves}\ for a discussion of the $K3$ cohomology lattice.} out of the $\Gamma^{3,19}$ lattice. This duality does not require a choice of $S^1_R$ but making such a 
choice will be useful for us later.

Note that the heterotic gauge-fields appear in M-theory by reducing $C_3$ on $H^2(S', \Z)$. Any unbroken non-abelian gauge symmetry in the heterotic string cannot be seen in supergravity but is correlated with an appropriate $ADE$ singularity of $S'$.  

We require a basic map of the parameters. The heterotic string is described by $(\alpha', \lambda_7, R)$ where $\lambda_7= e^{\phi_7}$ is the $7$-dimensional string constant defined 
in terms of the dilaton $\phi_7$, 
\begin{equation}
\phi_7 = \phi_{10} - \ff{1}{2} \log \frac{\vol(T^2\times S^1_R)}{\alpha'^{3/2}}. 
\end{equation}
 The M-theory compactification is characterized by $(\ell_p, E', V')$ where $\ell_p$ is the $11$-dimensional Planck constant. The map becomes particularly simple if we perturb $S'$ a little and assume that $\Pic(S')=2$. Comparing the masses of BPS particles in both compactifications then leads to the relations
\begin{eqnarray}
\alpha' & = & \ell_{p}^2, \nonumber\\
\lambda_7 &=& \left( \frac{V'}{\ell_{p}^4} \right)^{3/4}, \nonumber\\
R        & = & \ell_{p} V'^{1/2}E'^{-1}.
\end{eqnarray}
These relations can also be inverted:
\begin{eqnarray}
\ell_{p} &=& \sqrt{\alpha'}, \nonumber\\
E'        &=& \alpha'^{3/2} \lambda_7^{2/3} R^{-1},\nonumber\\
V'        &=& \alpha'^2 \lambda_7^{4/3}.
\end{eqnarray}
We have chosen to measure all our volumes in units of $\ell_{p}$.  To keep the resulting formulae uncluttered, we will subsequently set $\ell_{p} = 1$.

\subsubsection{The F-theory limit}

As before, let us suppose that the
Picard group of $S'$ is two-dimensional and generated by two null vectors $v,v^\ast$.  The K\"ahler form on $S'$ has the
general form
\begin{equation}
J = \alpha v +\beta v^\ast.
\end{equation}
Without loss of generality, we may suppose the fiber class is represented
by $v$, while that of the section by $v^\ast -v$, so that
\begin{eqnarray}
E' &=& J\cdot v = \beta,\nonumber\\
B' &=& J\cdot (v^\ast-v) = \alpha -\beta,
\end{eqnarray}
giving $J = (E'+B') v + E' v^\ast$.
Finally, the volume of $S'$ is also calibrated by $J$:
\begin{equation}
V' = \ff{1}{2} \int_{S'} J\wedge J = \ff{1}{2} J\cdot J = E' (E'+B').
\end{equation}

Now we can take the limit $R\rr \infty$ while holding fixed $\lambda_8 = \lambda_7 \sqrt{R}$ and $\alpha'$. In this limit we are left with heterotic on $T^2$. The elliptic fiber of $S'$ scales like 
$E' \sim R^{-4/3}$ while $B' \sim R^{2/3}$. In this limit M-theory goes over to type IIB string theory with a string scale $\alpha'_{IIB} \sim R^{2/3}$. With respect to this IIB scale, $B'$ has finite volume proportional to $\lambda_8$. This is the F-theory limit~\cite{Vafa:1996xn}.

\subsubsection{Reduction to three dimensions}
\label{reducetothree}

We can now further compactify both sides of this $7$-dimensional equivalence on an additional $K3$ surface $S$ to obtain a three-dimensional equivalence. Let us assume that $S$
is elliptically-fibered with section.  We denote the fiber and section volumes by $E$ and $B$, 
respectively.  This yields the dual description
\begin{eqnarray}
\text{Het. on}~S_H \times T^2\times S^1_R  &\leftrightarrow& \text{M-theory on}~S\times S', \nonumber\\
E_H, V_H, \lambda_3, R                 &\leftrightarrow& E, V,E',V'.
\end{eqnarray}
The relations between these parameters follow from the seven-dimensional parameter map 
described above and the relation between the seven-dimensional space-time metrics: 
\begin{equation}
G^7_{\text{Het.}}= V' G^7_{\text{M}}. 
\end{equation}
We find
\begin{eqnarray}
E_H & = & V' E, \nonumber\\
V_H & = & {V'}^2 V, \nonumber\\
R        & = & {V'}^{1/2} E'^{-1}, \nonumber\\
\lambda_3 & = & \lambda_7 V_H^{-1/2} = {V'}^{-1/4} V^{-1/2}.
\end{eqnarray}
These relations can also be inverted:
\begin{eqnarray}
E & = & E_H V_H^{-2/3} \lambda_3^{-4/3}, \nonumber\\
V & = & V_H^{-1/3} \lambda_3^{-8/3}, \nonumber\\
E' & = & V_H^{1/3} R^{-1} \lambda_3^{2/3}, \nonumber\\
V' & = & V_H^{2/3} \lambda_3^{4/3}.
\end{eqnarray}

The parameter map is clearly only part of the story. We have a choice of bundle/5-branes in the heterotic string satisfying~\C{hetbianchi}\ and a choice of flux/branes in M-theory satisfying~\C{Mtadpole}. These two choices are correlated. First, we note that the heterotic five-branes map directly to M2-branes.  The choice of heterotic gauge-bundle is interpretated as $G_4$-flux. A heterotic gauge-field arises from a class $\omega' \in H^2(S', \Z)$  while its field strength corresponds to a choice of class $ \omega \in H^2(S, \Z)$. So we can view the heterotic gauge-field in terms of M-theory  as 
\be
\frac{C_3 }{ 2\pi} = A \wedge \omega' 
\ee 
with field-strength
\be\label{hetbundle}
\frac{G_4 }{ 2\pi} = F_2 \wedge \omega'  = \omega \wedge \omega'. 
\ee

Lastly, we can write the parameter relations in terms of the $4$-dimensional heterotic coupling $\lambda_4^2 = R\lambda_3^2$ for later convenience,
\begin{eqnarray}
E & = & E_H V_H^{-2/3} \lambda_4^{-4/3} R^{2/3}, \nonumber\\
V & = & V_H^{-1/3} \lambda_4^{-8/3} R^{4/3}, \nonumber\\
E' & = & V_H^{1/3} R^{-4/3} \lambda_4^{2/3}, \nonumber\\
V' & = & V_H^{2/3} \lambda_4^{4/3} R^{-8/3}.
\end{eqnarray}
Taking $R\rr \infty$ is again the F-theory limit.

\subsection{A heterotic equivalence and implications for type IIA}
\label{hetequiv}

Now we want to use M-theory to connect two different heterotic theories. This duality will provide the heterotic analogue of the c-map. So we note that:
\begin{equation}
\begin{array}{ccccc}
\text{Het. on}~S_H \times T^2 \times S^1_R &\leftrightarrow& \text{M on}~ S\times S' &\leftrightarrow& 
\text{Het. on}~(T^2)' \times S^1_{R'} \times S'_H \\
\lambda_3, R, E_H, V_H, &\leftrightarrow& E,V,E',V' &\leftrightarrow& \lambda_3', R', E_H', V_H'. \nonumber
\end{array}
\end{equation}
The map works in the following way: we first trade the heterotic string on $S_H \times T^3$ for M-theory on $S \times S'$ where $S_H \rightarrow S$ and $T^3 \rightarrow S'$. The parameters $(E_H, V_H)$ for the $K3$ surface $S_H$ go over to $(E,V)$ for $S$. The parameters $(\lambda_3, R)$ determine $(E', V')$ for $S'$. This step uses the relations we just derived in section~\ref{reducetothree}. In M-theory both $S$ and $S'$ can be treated symmetrically. So we can construct a dual heterotic string theory by then sending $S \rightarrow (T^2)' \times S^1_{R'}$ and $S' \rightarrow S'_H$. The parameters $(E, V)$ of $S$ determine $(\lambda_3', R')$ in this dual heterotic compactification while $(E', V')$ determine $(E_H', V_H')$. 

Mapping the parameters along the lines of section~\ref{reducetothree}\ gives the relations, 
\begin{eqnarray} \label{hethetparam}
E_H' & = & R^{-1}\lambda_3^{-2} ,\nonumber\\
V_H' & = & \lambda_3^{-4}, \nonumber\\
R'     &  =&  E_H^{-1} V_H^{1/2}, \nonumber\\
\lambda_3' & = & V_H^{-1/4}.
\end{eqnarray}
For convenience, we also give $B'_H = V_H'/E_H' - E_H'$:
\begin{equation}\label{formulabase}
B_H' = (R-R^{-1}) \lambda_3^{-2}.
\end{equation}

Under this heterotic/heterotic duality, there is a natural action on the gauge bundle induced by exchanging the roles of $\omega$ and $\omega'$ in~\C{hetbundle}. Further,  note that the (dualized) vectors describing $T^2 \times S^1_R$ map under the duality to hypermultiplets describing the physics of $S'_H$. Therefore quantum corrections to the vector multiplets in one theory describe quantum corrections to hypermultiplets in the dual theory. In particular, small $\lambda_3$ maps to large $V_H'$ so perturbative loop corrections map to quantum corrections around large volume in the $(0,4)$ SCFT with target $S_H'$.  This is what we desire.  

This map therefore provides a means of determining quantum corrections to the dual $(0,4)$ SCFT. Namely, compute the tree-level and $1$-loop corrections to the vector moduli space metric in {\it three} dimensions, and re-intepret these corrections in the dual heterotic theory. These vector multiplet corrections have already been studied in four dimensions where the $1$-loop correction is known to have a rich structure~\cite{Harvey:1995fq, Marino:1998pg, Klemm:2005pd}. We expect the computation in three dimensions to be at least equally interesting. This computation will be examined elsewhere. 

What is missed in this approach are non-perturbative corrections to the vector moduli space like the monopole instantons mentioned in the introduction as well as NS5-brane corrections. We will describe a different limit that captures some of those corrections in the following section.  Toward that aim, let us re-write the parameter map~\C{hethetparam}\ in terms of the four-dimensional string coupling of the first
heterotic description. Recall that $\lambda_4^2 = R \lambda_3^2,$ so that
\begin{eqnarray}\label{fourrdimhethet}
E_H' & = & \lambda_4^{-2},\nonumber\\
V_H' & = & \lambda_4^{-4}R^{2},\nonumber\\
B_H' & = & (R^2-1) \lambda_4^{-2}.
\end{eqnarray}
We see that the limit $R\rr \infty$ with $\lambda_4$ fixed blows up the section of $S_H'$ leaving the volume of the elliptic fiber untouched. This is the limit in which perturbative computations in the heterotic string have already been explored.

However, there is a new limit suggested by~\C{fourrdimhethet}. Namely, taking $R\rr \infty$ while scaling  $\lambda_4$ to hold fixed the volume of cycles like  $V_H'$ ($\lambda_4 \sim \sqrt{R}$) or $B_H'$ ($\lambda_4 \sim R$). These are highly quantum limits of the $(0,4)$ SCFT because the elliptic fiber is shrinking to zero. Indeed a new dimension must emerge from these limits in the dual heterotic theory! The way this comes about can be seen by tracking the fate of KK modes on $S^1_R$ with
\be
M_n \sim \frac{n }{ R}. 
\ee
 These modes map to M2-branes wrapping $E'$ in M-theory. In turn, these M2-branes map to NS5-branes of the primed heterotic string wrapping $(T^2)' \times S^1_{R'}  \times E_H'$. This is a new quantum phenomena in the heterotic string: wrapped NS5-branes give rise to a new dimension.

While this limit of small fiber is unusual from the perspective of the perturbative sigma model, it is very natural from this duality. This is very much like the F-theory limit of $K3$ except studied in the {\it heterotic string}.  The reason this limit is so interesting is that we can use results from type II/heterotic duality to determine exact expressions for the vector multiplet couplings from curve counting in the type II string.

\vskip 0.5cm
\noindent {\em {\underline{Additional results from type IIA}}}
\vskip 0.5cm

For heterotic models with a type IIA dual description, we can attempt to learn more about our heterotic/heterotic duality from the type IIA dual; see~\cite{Aspinwall:1999xs}\ for earlier work in this spirit. To proceed, we need to determine the parameter map for the following equivalence: 
\begin{eqnarray}
\text{IIA on}~X \times S^1_r & \leftrightarrow & \text{Het on}~S_{\text{H}} \times T^2 \times S^1_R \nonumber\\
g_3, X_{\P^1}, r & \leftrightarrow & \lambda_3, B_H, E_H, R.
\end{eqnarray}
Note that $X$ is $K3$-fibered with section ${\P^1}$ of area $X_{\P^1}$. Matching space-time effective actions gives the relations, 
\begin{eqnarray}
%RIGHT:
R & = & g_3^{-1} r^{1/2} X^{-1/2}_{\P^1}, \nonumber\\
\lambda_3 & = & g_3^{1/2} r^{-1/4} X^{-1/4}_{\P^1}, \nonumber\\
B_H & = & g_3^{-2} r^{-1}, \nonumber\\
%WRONG, thanks ref.
%R & = & g_3^{-1} r^{3/2} X^{1/2}_{\P^1}, \nonumber\\
%\lambda_3 & = & g_3^{1/2} r^{-3/4} X^{-3/4}_{\P^1}, \nonumber\\
%B_H & = & g_3^{-2} r^{-1}, \nonumber\\
E_H &=& Y. 
\end{eqnarray}
The complex structure modulus $Y$ of $X$ is not distinguished except if there is an orientifold locus for $X$. If such a locus exists then at that locus $Y$ corresponds to the complex structure of the elliptic fiber. The inverse map is:
\begin{eqnarray}
%RIGHT:
g_3 &=& R^{-1/4} \lambda_3^{1/2} B_H^{-1/4},\nonumber\\
r      &=& R^{1/2} \lambda_3^{-1} B_H^{-1/2},\nonumber\\
X_{\P^1} &=& R^{-1} \lambda_3^{-2},\nonumber\\
%WRONG, thanks ref.
%g_3 &=& R^{-3/8} \lambda_3^{-1/4} B_H^{-3/8},\nonumber\\
%r      &=& R^{3/4} \lambda_3^{1/2} B_H^{-1/4},\nonumber\\
%X_{\P^1} &=& R^{-1} \lambda_3^{-2},\nonumber\\
Y      & = & E_H.
\end{eqnarray}
It is useful to re-write this in terms of the four-dimensional couplings:
\begin{eqnarray} \label{IIAfour}
%RIGHT:
g_4 &=& B_H^{-1/2},\nonumber\\
r      &=& \lambda_4^{-1} B_H^{-1/2} R,\nonumber\\
X_{\P^1} &=& \lambda_4^{-2},\nonumber\\
%WRONG, thanks ref.
%g_4 &=& B_H^{-1/2},\nonumber\\
%r      &=& \lambda_4^{1/2} B_H^{-1/4} R^{1/2},\nonumber\\
%X_{\P^1} &=& \lambda_4^{-2},\nonumber\\
Y      & = & E_H.
\end{eqnarray}
We can now see that the limit $R \rr \infty$ with $\lambda_4 \sim \sqrt{R}$ described at the end of section~\ref{hetequiv}\ sends $r\rr \infty$. It also sends the size of the base ${\P^1} \rr 0$ which is the opposite limit of the one usually studied in this duality. This limit, however, might be accessible in examples where the $\cF^{(g)}$ couplings have been directly computed in type IIA~\cite{Grimm:2007tm}. 

We are also free to scale $B_H$ as we wish since this is a hypermultiplet on the heterotic side so we can consider other scalings of $\lambda$ with $R$ and still ensure that $r \rr \infty$. Now in principle, this limit captures all quantum corrections to the hypermultiplet moduli space in the primed heterotic theory in the limit described at the end of section~\ref{hetequiv}. All such quantum corrections are encoded in the $\cF^{(g)}$ couplings expanded around small type IIA section ${\P^1} \rr 0$. 

\vskip 0.5cm
\noindent {\em  {\underline{Implications for type IIA}}}
\vskip 0.5cm

The final step is to ask what we learn about hypermultiplets in the type IIA theory dual to heterotic on $(T^2)' \times S^1_{R'} \times S'_H$. Let us denote the dual theory type IIA on $X' \times S^1_{r'}$. We expect $X'$ to correspond to the mirror of $X$~\cite{Perevalov:1998eg, Aspinwall:2005qw}. 

We have proposed two methods for determining quantum effects in the heterotic string on $S_H'$. The $1$-loop heterotic computation in three dimensions makes sense when the volume $V_H'$ is large. From~\C{formulabase}, we see that $B_H'$ can also be kept large. This is good news since this means we can see quantum corrections in IIA in the limit where $g_4'$ is small by using~\C{IIAfour}. We should therefore be able to distinguish between NS5-branes, sLags and perturbative corrections from the heterotic loop computation. 

The second proposal involves the use of the exact $\cF^{(g)}$ couplings in the IIA theory on $X$.  Once again the volume $V_H'$ can be kept large and fixed but in this limit $B_H' \sim R$ and so $g_4' \rr 0$. We would like to hold $g_4'$ finite (and small) with $r' \rr \infty$ so we should consider the scaling 
\be V_H' \sim 1/R^2. \ee 
This is a highly quantum limit in the heterotic string with the volume of the  elliptic fiber $E_H'$ becoming very small. The dual of the elliptic fiber $E_H'$ is a complex structure modulus $Y'$ of $X'$ so we are at least tuning the value of $Y'$. It is possible that this is a singular limit in complex structure moduli space in which the counting of brane instantons simplifies.  This needs to be explored further in specific examples.

\section{Curves and SLags in String Theory} \label{stringcurves}

In the preceding section, we described two specific proposals for computing classes of quantum corrections in heterotic and related dual  type IIA compactifications. We  
will now
restrict to type I/heterotic compactifications which admit an orientifold description~\cite{Sen:1997gv}. This will provide a more explicit construction which, at least in some examples, permits us to map all D-instantons in type I to brane instantons in M-theory or type IIA.  

\begin{figure}[bth!]
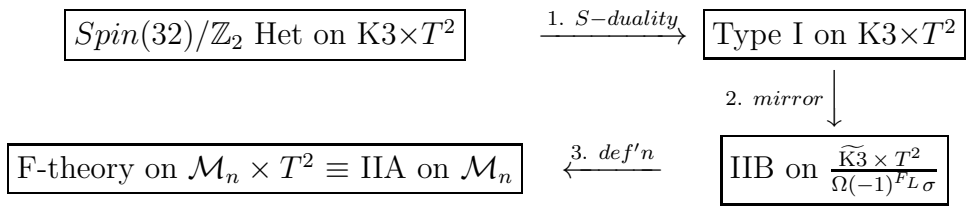

\[
\begin{CD}
\fbox{$Spin(32)/\ZZ_2$ Het on K3$\times T^2$} @>{1.\ S-duality}>>\fbox{Type I on K3$\times T^2$}  \\
%@>{}>> \fbox{Type I on K3$\times T^2$} \\ 
@.  @V{2.\ mirror}VV\\
 \fbox{F-theory on ${{\cal M}_n\times T^2}$ $\equiv$ IIA on ${\cal M}_n$}@<{3.\ def'n}<<   \fbox{\text{IIB on} $\frac{\text{$\widetilde{\text{K3}}\times T^2$ }}{\Omega(-1)^{F_L}\sig}$} \\
 \end{CD}
\]
\caption{The type IIA $\leftrightarrow$ heterotic duality sequence}
\label{fig:dualitymap}
\end{figure}

The duality chain we will exploit is illustrated in figure~\ref{fig:dualitymap}. In our subsequent discussion, we will investigate steps $2$ and $3$ in this duality sequence. We should emphasize that this particular duality chain is only valid for the $Spin(32)/\ZZ_2$ string. This is good enough since we want to explicitly track the fate of certain curves in $K3$.

\subsection{Type I $\leftrightarrow$ F-theory Duality}
\label{subsec:FIIA}

We start in six dimensions with the type I string on a $K3$ surface $S$. Equivalently, we can view this compactification as type IIB string theory on $S$ quotiented by world-sheet parity $\Omega$.\footnote{It is unfortunate that $\Omega$ is conventionally used to denote both world-sheet parity and holomorphic top forms. We hope the context will make the usage unambiguous.} 
 The gauge bundle
is characterized by one of the three choices of ${\widetilde w}_2$ given in~\C{nvscases}.  Each of these three cases is dual to F-theory compactified on an elliptically-fibered Calabi-Yau $3$-fold (with section) ${\cal M}_n$. The base of the elliptic fibration $B$ is a particular Hirzebruch surface $F_n$. 

More generally, any elliptically-fibered $3$-fold $\M$ with section can be used to construct an F-theory compactification. Such a space can be presented in Weierstrass form. Let $W$ be a  $\PP^2$ bundle over $B$ with homogeneous coordinates $(x,y,z)$ which we take to be sections of ${\mathcal O}(1)\otimes K^{-2}$, 
${\mathcal O}(1)\otimes K^{-3}$ and ${\mathcal O}(1)$,  respectively. The line-bundle
${\mathcal O}(1)$ is the degree one
bundle over the $\PP^2$ fibre and $K$ is the canonical bundle of $B$. Then
$\M$ is given by 
\be \label{weier}
 zy^2 = x^3+f xz^2+g z^3,
\ee
where $f$ and $g$ are sections of $K^{-4}$ and $K^{-6}$ respectively. There is a useful symmetry of $\M$ easily described in the presentation~\C{weier}; namely, inversion of the elliptic fiber which corresponds to the
$\ZZ_2$ action
\be\label{involution}
{\cal I}: y \rightarrow -y. 
\ee 
In turn, F-theory on ${\cal M} \times T^2$ is equivalent to type IIA string theory on ${\cal M}$. The additional torus factor does not affect the hypermultiplet moduli space because the moduli of the torus appear in vector multiplets.

This leads to our first conclusion. As described in section~\ref{reviewslag}, generic sLags in a $3$-fold do not have any particularly nice structure. However, sLags in type IIA or M-theory on $ {\cal M}$ which lift to Lorentz invariant six-dimensional instantons have a very special structure. If we denote the Poincar\'e dual to a particular sLag submanifold by $\alpha$ then this form must be odd under the involution~\C{involution},  
\be
{\cal I}^\ast( \alpha) = -\alpha. 
\ee  
As we explicitly demonstrated in the case of $T^6$, most sLags are not of this form. Since the hypermultiplet moduli space is unchanged as we decompactify the $T^2$ factor or equivalently take the F-theory limit, the contribution to the hypermultiplet couplings of sLags not of this form must cancel. This is true for any elliptically-fibered space $\M$. Said differently: the instanton sum in six dimensions must reproduce the result of the full sum over sLags in four dimensions. However, each instanton in six dimensions will map to a sLag with this special structure. This is a kind of localization phenomenon. 

Now let us return to the three cases of type I bundle appearing in~\C{nvscases}. These three choices are correlated with a choice of NS $B$-field~\cite{Sen:1997pm}. 
Recall that $H^2(K3,\ZZ)$ is an even self-dual lattice $\Gamma^{3,19}$ with signature $(3,19)$. In studying conformal field theory on $K3$, it is natural to add an extra copy of  $\Gamma^{1,1}$ and consider the lattice
\be
\Gamma^{4,20} = \Gamma^{1,1} \oplus 3\Gamma^{1,1}\oplus 2 (-\Gamma^{E_8}). 
\ee
This extra factor of $\Gamma^{1,1}$ is naturally associated to the 0-cycle and 4-cycle of the $K3$. Specifying a $B$-field is equivalent to choosing an element of $H^2(K3, \RR/\ZZ)$. A half-integral choice $B=\Lambda/2$ for some $\Lambda\in H^2(S',\ZZ)$ is compatible with the involution $\Omega$. In the resulting type I compactification, $\Lambda=\tilde{w}_2.$

We choose a mirror transform that trivializes the $B$-field and takes $S \rr S'$. Under this transform, 
\be
\Omega \stackrel{mirror}{\ra} \Omega (-1)^{F_L} \sig,
\ee
where $\sig$ is a quotient of $S$ by a $\ZZ_2$-symmetry. Such quotients were classified by Nikulin \cite{Nikulin} and are labeled by three integers $(r,a,\delta)$. It is important for us that $r$ is the rank of the sublattice $S^+\subset H^2(S',\ZZ)$ invariant under $\sig$. 

In~\cite{Sen:1997pm} it was determined that $S^+$ has rank $2$ and therefore $r=2$. Only for the case $\tilde{w}_2=0$ is $S^+\simeq \Gam^{1,1}$. The three classes of $Spin(32)/\ZZ_2$ bundle are mapped into the following quotients:
\bea \label{vbcases}
& (i)&  \tilde{w}_2=0 \leftrightarrow (r,a,\delta)=(2,0,0), \non \\
&(ii) & \tilde{w}_2 \cdot {\widetilde w}_2=0\ {\rm mod}\ 4 \leftrightarrow (r,a,\delta)=(2,2,0), \non \\
&(iii)& \tilde{w}_2 \cdot {\widetilde w}_2=2\ {\rm mod}\ 4 \leftrightarrow (r,a,\delta)=(2,2,1). \non 
\eea
For these cases, the fixed lattice $S^+$ has the metric
\be\label{invmetrics}
(i) \, \bpm 0 & 1 \\ 1 & 0 \epm, \qquad (ii) \, \bpm 0 & 2 \\ 2 & 0 \epm, \qquad (iii) \, \bpm 0 & 2 \\ 2 & 2 \epm.
\ee
This result does not require the division of the moduli space of $S$ or $S'$ into K\"ahler and complex structure moduli; however, it is useful to note that $S^+ = \Pic (S')$ for generic choices of complex structure compatible with $\sigma$. Associated to these Nikulin quotients are $3$-folds ${\cal M}_n$ with $B$ given by $F_4, F_0$ and $F_1$, respectively. At a special locus in moduli space, these  ${\cal M}_n$ $3$-folds can be realized as Borcea-Voisin $3$-folds~\cite{Morrison:1996pp}\ described  generally in section~\ref{sLagBV}.

\subsection{The $K3$ Mirror Map}

To understand the mapping of a curve in $S$ into a sLag in ${\cal M}_n$, we will need some details regarding mirror symmetry for $K3$. We want to describe the mirror transform between type I on $S$ and the F-theory type IIB orientifold of $S'$. Background information can be drawn from \cite{Aspinwall:1994rg, Aspinwall:1996mn, gross-1996, Huybrechts:2002ap}. 

First consider a $K3$ surface with cohomology lattice
\bea
H^2(K3,\ZZ)&\simeq& L \non \\
&=&M\oplus T \non \\
&=&M \oplus P\oplus \check{M}.  \label{latticedecomp}
\eea
The lattice $T$ is the transcendental lattice and contains all classes in $L$ orthogonal to the K\"ahler form. We have distinguished a $\Gamma^{1,1}$ factor $P$ from $T$ which corresponds to a choice of sLag fibration with a torus fiber. The lattice $M = \Pic(K3)$. This decomposition into $M$ and $T$ depends on the choice of complex structure. The quantum lattice is formed by adjoining another copy of $\Gam^{1,1}$  for the zero and four cycles: 
\be
H^{2}_{qu}(K3,\ZZ)\simeq Q \oplus L.
\ee

The mirror map is an identification of geometrically distinct $K3$ surfaces which give rise to the same $(4,4)$ SCFT. Each curve holomorphic in some complex structure of $S$ can support a BPS D-instanton. What we would like to show is that each of these curves maps to a brane configuration in the mirror $S'$ which lifts to a BPS instanton of type IIA on ${\cal M}_n$.

The map is formulated at the level of covering spaces of the physical moduli space. The covering space of the complex structure moduli space is the period domain $D_M$. Choosing a point in this covering space corresponds to choosing a two-form $\Om \in H^2(K3, \CC)$ satisfying
\be
 \Om \cdot \Om =0,\quad \Om\cdot \Ombar >0. 
\ee
The choice of $\Om$ is unique up to a phase. By definition $\Om$ lies in a complexified subspace spanned by $T \in H^2_{qu}(K3,\ZZ)$. 
The covering space of the complexified K\"ahler moduli space is called the tube domain,
\be
T_{M} =\{ B+i\om \in M\otimes_\ZZ \CC | \ \om \cdot \om >0 \}
\ee
where $B\in H^2(K3,\RR/\ZZ)$ is defined modulo integral shifts. With suitable definitions~\cite{Huybrechts:2002ap, Dolgachev}, the mirror map corresponds to the exchange 
\be\label{defmirrormap}
\Om \, \leftrightarrow \, e^{B+i \om}. 
\ee
We will demand that our sLag fibration $P$ satisfy $\Lambda \cdot P =0$ then mirror symmetry will exchange $Q$ and $P$ in a simple fashion~\cite{Sen:1997pm}. Choose primitive isotropic vectors $E_1,E_2\in Q$ so that $E_1\cdot E_2=1$ and $E_1\cdot E_1= E_2 \cdot E_2=0$ then the maps
\bea
&& \phi_1:T_{M}(S) \ra D_M(S')  \non \\
&& \phi_2:D_{M}(S) \ra T_M(S')  \non 
\eea
are given by 
\bea \label{mirrorphi}
\phi_1(B+i \om) = \Om' &=& \left( B+ E_2 + \half(\om \cdot \om - B \cdot B) E_1 \right) \non\\
&& + i \left( \om - (\om \cdot B) E_1 \right), \non\\ 
\phi_2(\Om) = B' + i\om' &=&  \pr( \Om) - (\Om \cdot B) E_1, 
\eea
where the projection $\pr:P\oplus \check{M}\ra \check{M}$. These formulae essentially encode the relation~\C{defmirrormap}. As a final comment, note that
for the case where $B$ is purely a $(1,1)$ form, the map exchanges the quantum Picard lattice with the transcendental
lattice as illustrated in figure~\ref{fig:lattice}~\cite{Aspinwall:1994rg}. 

Now we start on $S$ with $B=\Lambda/2$ and perform a mirror transform to $S'$ that exchanges $Q$ with $P$. We want the resulting space to have no $B$-field so we impose the conditions
\be
\Re\left( \pr( \Om) \right) = 0, \qquad \Re (\Om \cdot \Lambda/2) =0 . 
\ee
We are free to tune the complex structure moduli of $S$ so these conditions are satisfied. 

\begin{figure}\
%[bth!] 
\begin{center}
\begin{tabular}{|c|c|c|}
\hline
& $S$ & $S'$ \\
\hline
Transcendental &$P\oplus \check{M}$& $Q\oplus M$ \\
\hline
Quantum Picard& $Q\oplus M$ &$P\oplus \check{M}$ \\
\hline
\end{tabular}
\end{center}
\caption{The exchange of lattices for the case of $(1,1)$ $B$-field.}
\label{fig:lattice}
\end{figure}
\vskip 0.5cm
\noindent {\em (i) {\underline{No $B$-field}}}
\vskip 0.5cm

We start with the simplest case where there is no $B$-field on $S$ and we have imposed the condition that $B'$ vanish on $S'$. In complex structure $I$, we denote 
holomorphic two-form and K\"ahler form of $S'$ by
\bea
\Omega_I' =\Re \Om_I' + i \Im \Om_I' &=& \phi_1(i \om) \non \\
&=& E_2 + \half(\om \cdot \om) E_1 + i \om, \\
i\om_I' = \phi_2(\Om) &=& \pr( \Om) .
\eea
We see that $\Omega_I' \in T'=Q \oplus M$ and is orthogonal to $P \oplus\check{M}$. So we note that
\be
\om_I' \in  \check{M}, \qquad
\Re \Om_I' \in Q, \qquad
\Im \Om_I' \in  M,
\ee
and thus in this complex structure, we see that  $E_1,E_2 \in Q$ are both sLags with zero phase. 

Now in this case
with no $B$-field, the invariant lattice $S^+$ is identified with $Q$~\cite{Sen:1997pm}.  So we make an $SU(2)$ rotation to complex structure $K$ chosen so that
\be\label{compK}
\om_K' = \Re \Om_I', \qquad
\Re \Om_K' = \Im \Om_I', \qquad
\Im \Om_K' =\om_I'.  
\ee
With this choice, all elements in $M$ and $\check{M}$ are orthogonal to $\om_K'$ and therefore admit Lagrangian representatives. As discussed in section  
\ref{sec:sLagK3}, any cycle admitting a Lagrangian representative in $K3$ admits a sLag representative for some choice of phase. 

This choice of complex structure is also compatible with a holomorphic action for the Nikulin quotient on $S$, discussed in section~\ref{subsec:FIIA}, which acts by sending
\be\label{niKK}
\Omega_K'\ra -\Om_K'.
\ee
In this frame, $E_1, E_2 \in S^+$ so, following the discussion in  section~\ref{sLagBV}, all elements of the transcendental lattice of $S_K'$ 
will lift to sLags in the corresponding Borcea-Voisin three-fold. As we will discuss further in section~\ref{SubHetIIA}, this is in agreement with our basic expectations about the mapping of BPS D-instantons.

\vskip 0.5cm
\noindent {\em (ii) {\underline{With $B$-field}}}
\vskip 0.5cm

Now we wish to turn on a discrete $B$-field on $S$ and see how this alters the previous conclusions. Again we assume that there is no $B$-field on $S'$. So consider $B = \Lambda/2$ on $S$ where we assume that $\Lambda \cdot P =0$. In complex structure $I$, we see that
\bea
\Omega_I' 
&=& \Lambda/2+ E_2 + \half(\om \cdot \om - \Lambda^2/4) E_1 
\non \\
&& + i \left( \om - (\om \cdot \Lambda) E_1/2 \right) , \non\\
i\om_I' &=& \pr( \Om) - (\Om \cdot \Lambda/2) E_1 .
\eea
As before, we rotate to complex structure $K$ given in~\C{compK}. In this complex structure, we see that $\Pic(S_K')$ is generated by
\be\label{rotatedpic}
\left\{ E_1, 2E_2 + \Lambda\right\}
\ee
since these two lattice vectors are orthogonal to  $\Im \Om_I'$ and $\om_I'$. The metric on the Picard lattice $\Pic(S_K')$
is readily computed from~\C{rotatedpic}, 
\be
\bpm  0 & 2 \\ 2 & \Lambda^2  \epm. 
\ee
Comparing these metrics with~\C{invmetrics}, we identify $\Pic(S_K')$ with the fixed lattice $S^+$. So with this choice, the Nikulin quotient acts holomorphically with the action given in~\C{niKK}. The generators given in~\C{rotatedpic}\ are the mirrors of the generators for $S^+$ given in~\cite{Sen:1997pm}. 

So again following the discussion in section~\ref{sLagBV}, we can conclude that all elements of the transcendental lattice of $S_K'$ lift to sLags of the associated Borcea-Voisin three-fold. It is important to note that curves in the homology class $[\Lambda]$ are not in the transcendental lattice. This is a reflection in the mirror of the observation~\cite{Witten:1999eg}\ that $KO(S)$ and $KO_{\widetilde{w_2}}(S)$ are different precisely in the direction $\Lambda$. This appears to be how the IIB orientifold mirror realizes the wrapping obstruction induced by the ``no vector structure'' condition. 

However the combination, 
\be
[\Lambda] - \frac{ \Lambda^2 }{ 2} [E_1],
\ee
is part of the transcendental lattice. A brane in this homology class in $S$ is initially a bound state of $D1$-brane and $D(-1)$-brane charge. However in type I string theory, there are no BPS $D(-1)$-branes so this configuration is not BPS.  It would be interesting to understand more deeply how this BPS constraint is realized in type IIA directly by studying the spectrum of BPS brane configurations on these different Borcea-Voisin spaces. There should be a significant difference between the cases with $a \neq 0$ and $a=0$ appearing in~\C{vbcases}. 

\subsection{Type IIA duals of distinguished submanifolds}
\label{SubHetIIA}

We now describe how different distinguished type I $D1$-instantons wrapping curves in $H^2(S,\ZZ)$ are mapped into the type IIA theory on ${\cal M}_n$.
For our purposes, the sLag fibered $K3$ surface $S$ has three distinct homology classes: the first corresponding to the section of the 
fibration; the second being the class of the generic fiber; and the third class consisting of 
all the singular fibers.   We will now discuss these cases in detail.

\vskip 0.5cm
\noindent {\em (i) {\underline{The section $S^2$}}}
\vskip 0.5cm

The choice of  sLag fibration corresponds to the choice of $P$ in~\C{latticedecomp}. Let $(P_1, P_2)$ be a basis for $P$
with
\be
P_1 \cdot P_1 = -2, \quad P_2 \cdot P_2 =0, \quad P_1 \cdot P_2=1. 
\ee 
We choose $P_1$ to be the class of the section. Under the mirror transform, a $D1$-brane in $P$ maps to a combination of wrapped $0$-branes and $4$-branes. A D-brane configuration with pure $D4$-brane charge lifts to a Euclidean NS5-brane in type IIA on ${\cal M}_n$. So any $D1$-brane wrapping a submanifold in a homology class with non-zero projection along $P_1$ will map to an instanton in type IIA with NS5-brane charge.        

\vskip 0.5cm
\noindent {\em (ii) {\underline{The generic fiber}}}
\vskip 0.5cm

The charge associated to the elliptic fiber $P_2$ is more subtle. In the mirror, we need to understand the lift of $D(-1)$ instanton charge to M-theory/type IIA on ${\cal M}_n$. This is not well understood. In simpler cases, we can provide a picture for how $D(-1)$ instanton effects are realized in M-theory. A particularly well studied case is M-theory on $T^2$ which is dual to type IIB on $S^1$. In this case,  $D(-1)$ instantons renormalize higher derivative gravitational couplings like the $R^4$ terms in the IIB string effective action. However in M-theory, these effects are reproduced by summing over Kaluza-Klein modes of $T^2$~\cite{Green:1997as, Green:1999qt}. So we expect the effects of these $D(-1)$ instanton configurations in F-theory to be reproduced by perturbative effects in type IIA on ${\cal M}_n$. 

\vskip 0.5cm
\noindent {\em (iii) {\underline{Singular fibers}}}
\vskip 0.5cm

All other curves in $S$ come from singular fibers and generically include curves of any genus. The self-intersection number of the curve is invariant under the mirror symmetry. Let us consider a curve in a class that maps to the transcendental lattice of $S'$. Each such curve lifts to a sLag in ${\cal M}_n$. The sLag is constructed from the curve in a manner described in section~\ref{sLagBV}.  In section~\ref{topology}, we will use physical arguments to conjecture  a relation between the topology of a given curve and that of its image sLag: namely, that a curve with genus $g$ maps to a sLag $L$ with $b_1(L) \geq g$.  
 
There is a particularly interesting sLag in a Calabi-Yau $3$-fold whose existence in general was conjectured by SYZ~\cite{Strominger:1996it}\ and constructed in the Borcea-Voisin cases in~\cite{gross-1996}.  This construction involves a sLag fibration of $S'$ in complex structure $K$ given in~\C{compK}\ where $\Pic(S_K') = S^+$. These sLags in combination with a sLag from $T^2$ becomes a $T^3$ and an $S^3$ of the $3$-fold~\cite{gross-1996}.  It would be interesting to extend this analysis to other sLags originating from curves in $S$; in particular, one could check the relation $b_1(L) \geq g$.  The sLag fibration of $S'$ used in~\cite{gross-1996}\ is mirror to a sLag fibration of $S$ distinct from the one we chose for mirror symmetry.

 \section{F-terms in $\N=2$ Supergravity From String Theory}
 \label{fterms}

Four-dimensional $\N=2$ supergravity admits higher derivative F-term couplings constructed
purely from either vector multiplets or  hypermultiplets. The former are sometimes called
untwisted F-terms while the latter are denoted twisted F-terms. The vector multiplet F-terms satisfy a known recursion relation~\cite{Antoniadis:1995zn, Bershadsky:1993cx}. It seems likely that space-time supersymmetry together with duality implies these relations along the lines sketched in~\cite{Sethi:2004np}. If true, we expect similar relations to hold for the twisted hypermultiplet F-terms.

What we can observe is that the string coupling often sits purely in a vector multiplet or a hypermultiplet. If this is true, 
we can expect these BPS space-time couplings to enjoy special renormalization properties when computed on the string
world-sheet.  In this section, we will examine the perturbative renormalization properties of these F-terms when computed in type IIA, heterotic or type I  
string theory. Both for completeness and contrast, we will begin by summarizing the structure of vector multiplet F-terms before turning to the hypermultiplets.

 %%%%%%%%%%%%%%%%%%%%%%%%%%%%

  \subsection{Vector multiplet F-terms}

One approach to constructing F-terms in four-dimensional $\N=2$ Poincar\'e supergravity is by gauge-fixing chiral couplings in  $\N=2$ conformal supergravity. For a nice review of this approach see~\cite{Mohaupt:2000mj}. The action of $\N=2$ conformal supergravity contains the chiral terms

\bea
&& \cL_{F}=\sum_{g\ge 0}  \cL_{F}^{(g)}, \\
&& \cL_F^{(g)} = \int \! d^4x\! \! \int \! d^2 \theta d^2 \ttha\, \W^{2g} F^{(g)}(X^I(z)),\quad I=0,\ldots,n.
\eea
Here $\W$ is the Weyl multiplet and $X^I(z)$ are $n+1$ vector multiplets  which transform as a section of 
an $Sp(2n+2)$ bundle over the moduli space ${\cal M}_V$. On gauge-fixing to obtain Poincar\'e supergravity, the Planck mass is introduced via  
the gauge choice
\be
X^I=m_{pl}e^{K(z,\zbar)/2}X^I(z),
\ee
where $K(z,\zbar)$ is the K\"ahler potential, and it is convenient to use special coordinates
\be
z^0=1,\ \ z^A=X^A/X^0,\ A=1,\ldots,n.
\ee
In components, we obtain the F-term couplings

\be \label{Lcomp}
\cL_{F}^{(g)}\sim \kappa^{2g-2} \int \! \! d^4 x R^2 T^{2g-2} e^{(1-g)K(z,\zbar)} \cF^{(g)}(z^A)+\ldots
\ee
where $R$ is the anti-self-dual projection of the Riemann tensor, $T$ is the anti-self-dual graviphoton field strength and

\be
\cF^{(g)}(z^1,\ldots,z^n)=F^{(g)}(1,X^1/X^0,\ldots,X^n/X^0).
\ee
The omitted terms in~\eq{Lcomp}\ constitute the supersymmetric completion. 
The computation of these low energy couplings from string theory is quite different in type IIA, heterotic and type I string theory. 
%
%%%%%%%%%%%%%%%%%%%%%%%%%%%%%

\vskip 0.5cm
\noindent {\em {\underline{Vector F-terms in type IIA string theory}}}
\vskip 0.5cm

In type IIA on $X$, the string coupling is part of a hypermultiplet while vector multiplets describe K\"ahler moduli of the compactification space $X$. Therefore, the only $g_s$ dependence in space-time vector couplings is via the gravitational coupling: $\kappa\sim g_s$. This implies that \eq{Lcomp} is computed exactly by a $g$-loop world-sheet calculation. Indeed, the superstring calculation of this coupling reduces to the genus $g$ topological A-model partition function \cite{Antoniadis:1993ze, Bershadsky:1993cx}. 

That that this coupling {\it only} receives contributions at genus $g$ can also be seen directly from the world-sheet. At genus $g$, we insert $2g-2$ graviphoton vertex operators in the $(-1/2)$ picture and two gravitons in the $(0)$ picture. To cancel the ghost charge from the curvature of the Riemann surface as well as the graviphoton insertions, we must also insert $3(g-1)$ picture changing operators%
\footnote{Recall that the $\beta\!-\!\gam\ U(1)$ ghost charge is saturated by inserting $n_{\cP}$ picture changing operators where 
\be n_{\cP}=2g-2+n_{NS}+n_{R}/2 \non \ee  and $n_{NS},n_{R}$ are the number of insertions in the $(-1)$ and $(-1/2)$ picture, respectively.}. 
With these insertions, the vanishing of the total $U(1)_R$ charge is satisfied only by terms in the world-sheet correlator which include the $e^{\phi}G^-$ part of the picture changing operators. 

If we were to consider computing~\eq{Lcomp} at genus $g'<g$ then the $U(1)_R$ charge condition could not be saturated since the graviphoton insertions still have $U(1)_R$ charge $3(g-1)$. This charge cannot be canceled by the $2(g'-1)+(g-1)$ picture changing operators and therefore the correlator vanishes.   

To understand the case with $g'>g$, we decompose the picture changing operator $\cP$ into an internal component and a space-time component

\be
\cP=\cP_{st}+\cP_{int}=e^{-\phi}G_{st}+e^{-\phi}G_{int}.
\ee
To saturate the internal $U(1)_R$ charge condition requires that half of the additional $2(g'-g)$ picture changing operators contribute either via  their $e^{\phi}G^{+}_{int}$ term or via their $e^{\phi}G^{+}_{st}$ term. In the former case, $e^{\phi}G^{+}_{int}$ annihilates the chiral graviphoton vertex operators. In the latter case, we must bring down more factors of space-time momentum which increases the derivative order of the space-time coupling. So either way, this particular coupling vanishes. 

%
%%%%%%%%%%%%%%%%%%%%%%%%%%%%
\vskip 0.5cm
\noindent {\em {\underline{Vector F-terms in heterotic string theory}}}
\vskip 0.5cm

In heterotic on $K3 \times T^2$, the string coupling is part of a vector multiplet.  In this case, the Peccei-Quinn symmetry is a powerful constraint on the axion-dilaton dependence. For example, the K\"ahler potential has the form~\cite{Antoniadis:1995zn}  

\be
e^{K(z,\zbar)} = g_s^2+\ldots,
\ee
where omitted terms are non-perturbative in the string coupling. This implies that the couplings~\eq{Lcomp} receive perturbative contributions from $1$-loop only. We should point out are that there are also tree-level contributions to $\cF^{(0)}$ and $\cF^{(1)}$ which arise because $g=0$ and $g=1$ surfaces admit conformal killing vectors. This same caveat applies throughout our discussion.   

This conclusion can also be reached by directly studying the heterotic world-sheet. This is again because at higher genus, some of the additional needed picture changing operators must contribute $e^{\tphi}\tilde{G}^+$ factors which annihilate the chiral insertions. So again, it is the special structure of the space-time F-terms which leads to restrictions on contributions from string loops.
%%%%%%%%%%%%%%%%%%%%%%%%%%%%
%

\vskip 0.5cm
\noindent {\em {\underline{Vector F-terms in type I string theory}}}
\vskip 0.5cm

In type I on $K3$, the string coupling resides in a hypermultiplet. However on $K3 \times T^2$, the resulting four-dimensional string coupling does not sit cleanly in either a vector multiplet or a hypermultiplet. It is easy to see that this must be the case: on the one hand, hypermultiplets are renormalized by $D1$-instantons. On the other hand, vectors are renormalized by $D5$-brane instantons. Therefore both flavors of multiplet detect the value of the string coupling. 

Computations of vector F-terms in type I on $K3 \times T^2$ were studied in~\cite{Serone:1996bk, Morales:1997zv}. Here we just summarize one particular point which is useful for us later. The space-time SUSY operators are
\bea
&& Q_{\al,1}^{(L)}+Q_{\al,1}^{(R)} = \oint dz e^{-\phi/2} S_{\al} e^{iH_T/2} \Sig + \oint d\zbar e^{-\tphi/2} \tS_{\al} e^{i\tH_T/2} \widetilde{\Sig},  \\
&& Q_{\al,2}^{(L)}+Q_{\al,2}^{(R)} = \oint dz e^{-\phi/2} S_{\al} e^{iH_T/2} \ol{\Sig} + \oint d\zbar e^{-\tphi/2} \tS_{\al} e^{i\tH_T/2} \ol{\tSig}.
 \eea
The graviphoton vertex operator is obtained by spectral flow from the graviton vertex operator:
\be
V_{gp}=(Q_{\al,1}^{(L)}+Q_{\al,1}^{(R)} )(Q_{\al,2}^{(L)}+Q_{\al,2}^{(R)}) V_g.
\ee

There are two different types of terms which arise from the two spectral flows acting on the graviton. We could consider 
$(Q_L Q_L+Q_R Q_R)V_{g}$ type terms or we could consider $(Q_L Q_R + Q_R Q_L)V_{g}$. Roughly speaking, these two sets of terms 
look respectively like the graviphoton from the heterotic string on $K3\times T^2$ and the graviphoton from type II on $X$. However as we will see, the F-terms are
all generated at only one loop. 

The picture changing operator is the world-sheet supercharge, 
\bea
\cP&=&e^{\phi} G  =e^{\phi}(G^++G^-) \non \\
&=&e^{\phi} \left(\del \Xbar_1 \psi_1+\del \Xbar_2 \psi_2 + \del \Xbar_{T} \psi_T + e^{i\frac{ H_{K3}}{\sqrt{2}}} \ol{\widehat{G}}_{K3}\right) + \non \\
&&   e^{\phi} \left(\del X_1 \psibar_1+\del X_2 \psibar_2 + \del X_{T} \psibar_T + e^{-i \frac{H_{K3}}{\sqrt{2}}} \widehat{G}_{K3}\right),
\eea
where $(X_1, X_2)$ denote the space-time coordinates while $X_T$ denotes the $T^2$ coordinate. The operator $\widehat{G}_{K3}$ is the neutral piece of the $K3$ supercurrent.
 To find which string loops can generate these F-terms, we must balance the various $U(1)$ charges on the world-sheet. 
 
 Schematically, we see that the graviphoton vertex operators have zero $U(1)_R$ charge from the $K3$ CFT, $(+1)$ $U(1)_R$ charge from the $T^2$ CFT and $(-1)$ 
total charge from the left-moving and right-moving ghost $U(1)$ combined.  One way to proceed is to observe that the PCO has $T^2$ $U(1)_R$ charge $(\pm 1)$ and $U(1)$ ghost charge $(+1)$. 

So to cancel the $U(1)$ charge of the $T^2$ factor, we would need to insert $(2g-2)$ PCO's which could only contribute through their $e^{\phi}G^-_{T^2}$ factors. By doing this we see that this condition of canceling the $T^2$ charge has resulted in zero net ghost charge; thus this amplitude is only non-zero at 1-loop.

%%%%%%%%%%%%%%%%%%%%%%%%%%%%
%%%%%%%%%%%%%%%%%%%%%%%%%%%%
\subsection{Hypermultiplet F-terms} \label{ss:Hypermultiplets}

The hypermultiplet F-terms have proven more difficult to characterize in either conformal or Poincar\'e supergravity; see~\cite{Rocek:2005ij, deWit:2001dj, Neitzke:2007ke, Robles-Llana:2006ez, Anguelova:2004sj} for recent progress. For our needs, a component formulation is sufficient. Let us recall that in certain limits of the 
moduli space ${\cal M}_H$, there is a distinguished universal hypermultiplet $q_Z$. The proposed F-terms are constructed as follows: decompose $q_Z$ into two complex scalars $\phi_S$ and $\phi_Z$. Such a decomposition only makes sense locally on the moduli space. In terms of $(\phi_S, \phi_Z)$, we consider the couplings

\bea
&& \cL_{\tF}=\sum_{g\ge 0} \cL_{\tF}^{(g)}, \label{eq:hypercoup} \\
&& \cL^{(g)}_{\tF}=\kappa^{2g-2}\int d^4 x \sqrt{g}  (\del_\mu \del_\nu \phi_S)(\del^\mu \del^\nu \phi_S) (\del_\mu \phi_Z\del^\mu \phi_Z)^{g-1} \tF^{(g)}(q^i); \quad  g \geq	 1. \non
\eea
The $q^i$ are the remaining hypermultiplets and $\cL^{(0)}_{\tF}$ contains the hypermultiplet kinetic terms. These couplings treat $\phi_S$ and $\phi_Z$ asymmetrically. A more natural formulation of these twisted F-terms should not involve such a distinction. A superspace completion of these higher derivative terms has been proposed in~\cite{Rocek:2005ij}. 

In the theory with $\cN = 4$ supersymmetry, the $\int d^4x (\p\p S)^2$ coupling is related by supersymmetry to $\int d^4x R^2$.  Both couplings have been computed by a one-loop computation in type IIA---the former in~\cite{Antoniadis:1997zt} and the latter in~\cite{Harvey:1995fq}.

%%%%%%%%%%%%%%%%%%%%%%%%%%%%
\vskip 0.5cm
\noindent {\em {\underline{Type IIA string theory}}}
\vskip 0.5cm

The type IIA world-sheet  CFT computation of \eq{eq:hypercoup} is very similar to the computation of \eq{Lcomp}. Despite the fact that the string coupling is part of a hypermultiplet, the $\cL_{\tF}^{(g)}$ are generated only at $g$-loops in perturbation theory. These couplings are given by the topological B-model genus $g$ partition function~\cite{Antoniadis:1993ze}. In addition, there are all the interesting sLag and NS5-brane instanton effects which are non-perturbative in the string coupling.

%%%%%%%%%%%%%%%%%%%%%%%%%%%%
\subsubsection {Heterotic string theory}
\label{ftermsheterotic}
This class of couplings has not been previously considered from the world-sheet perspective. The only $g_s$ dependence is through  the gravitational coupling $\kappa\sim g_s$ since the string coupling sits in a vector multiplet.  Consequently, we expect the amplitudes which determine this coupling to be $g$-loop only. We will demonstrate this directly from the string world-sheet.

The universal hypermultiplet in the heterotic string comes from reducing the NS $B$-field on the $3$ anti-self-dual forms of $K3$ together with the overall volume. The corresponding vertex operators for the hypermultiplet scalars $\phi_S$ and $\phi_Z$ in the $(-1)$ picture are

\bea
&& \cV^{(-1)}_S=J_{i\bar{\jmath}} e^{-\phi}\!\! :\!\! \del X^i \tpsi^{\bar{\jmath}} e^{ip\cdot X}\!:\!\! (z,\tz), \\
&& \cV^{(-1)}_Z=\Om_{ij}e^{-\phi}\!\!:\!\! \del X^i \tpsi^{j} e^{ip\cdot X}\!:\!\!(z,\tz),
\eea
where $J_{i\bar{\jmath}} $ is the K\"ahler form and $\Om_{ij}$ is the holomorphic $2$-form of $K3$. These vertex operators are transformed into each other by two spectral flows. 

The coupling 
\be \label{ddsdds}
\int d^4 x \sqrt{g} (\del \del \phi_S)^2 \tF^{(1)}(q^i)
\ee 
is generated at genus $1$ only. At genus $1$, there is no $U(1)$ background charge so the $U(1)$ charges of the two insertions of $S$ must cancel amongst themselves. The two point function of $S$ vanishes on shell so to compute~\eq{ddsdds}, we must consider the four point function of two gravitons and two $S$ insertions which is of order $O(p^4)$.

The key difference between the heterotic and type IIA calculations is that on the heterotic side, $\phi_Z$ is an NS-NS field while it is an R-R field in type IIA. In terms of calculating space-time effective actions from the world-sheet, recall that Ramond insertions compute field strengths in space time while NS insertions compute potentials. To see that \eq{eq:hypercoup} is a genus $g$ amplitude, we again use $U(1)$ charges. 

Consider  \eq{eq:hypercoup} computed at genus $g'$. The two insertions of $\cV_{S}$ are in the zero picture and must be accompanied by two gravitons just as in the genus $1$ case. This alone cancels all $U(1)$ charges and so we can consider the $\cV_{Z}$ insertions separately. Each insertion of $\cP_{st}$ provides a power of space-time momentum from its contraction with $e^{ip X}$. Since we need $2g-2$ powers of momentum from the $\phi_Z$ insertions, we see that $n_{st}=2g-2$. Each insertion of $\cP_{int}$ will provide $(-1)$ units of internal $U(1)_R$ charge. Further, the cancellation of the $\beta\!-\!\gamma$ ghost number implies that

\be
n_{int}+n_{st}\equiv n_{\cP}=2g'-2+2g-2,
\ee
and the $U(1)_R$ cancellation gives 

\be
n_{int}=n_{NS}=2g-2.
\ee
From this we see that $g=g'$.

%%%%%%%%%%%%%%%%%%%%%%%%%%%%
\subsubsection{Type I string theory} 

The hypermultiplet F-term calculation in the type I string on $K3\times T^2$ is quite interesting. The basic difference in the world-sheet calculations of heterotic versus type I is that the $\phi_S$-field is related to the string coupling in type I. Indeed in type I on $K3$, this scalar is the dilaton-axion and is therefore the identity in the chiral ring of the internal CFT. This is quite different from the heterotic string where the corresponding vertex operator is non-trivial in the internal CFT. This really should be the case because the heterotic string CFT computation is exact while the $D1$-instantons are missed in type I perturbation theory. 

It is interesting to note that by comparing the type I calculation with the type IIA calculation, we can conclude that the type I world-sheet calculation is equivalent to a reformulation of the topological B-model on the dual space $X$. It would be fascinating if this observation could lead to some practical alternative to the traditional B-model calculations.

The $\phi_Z$-field vertex operator is
\be
V_{Z}=(\Qbar_{\dal,2}^{(L)}+\Qbar_{\dal,2}^{(R)} )(Q_{\al,1}^{(L)}+Q_{\al,1}^{(R)}) V_S.
\ee
The main difference between this vertex operator and the graviphoton vertex operator is that this one has vanishing $U(1)$ charge from the $T^2$ and (left plus right-moving) $(+2)$ units of $U(1)_R$ charge from the $K3$ CFT. 

The picture changing operator is
\be
\P=e^{\phi} (G^+ + G^-) + e^{\tphi} (\tG^+ + \tG^-).
\ee
If we insert $2g-2$ copies of $V_Z$ then since $V_Z$ has four terms, there could be many terms which  in principle could contribute. Each of these four terms has a different amount of left and right-moving charge for both the ghost $U(1)$ and $U(1)_{K3}$.  However the sum of left and right-moving charges is equal for each term. Note that the $U(1)_{T^2}$ charge of $V_Z$ vanishes. Likewise for multiple insertions of $\P$, there are many terms all of which have a common left plus right sum of charges.

Now recall that $G^\pm$ has $\pm1$ $U(1)$ charge and $\Sig_{K3}$ has charge $1$. So if we insert $2g-2$ copies of $V_Z$ then a possible term 
is
\be
\left( \Qbar_{\dal,2}^{(L)}Q_{\al,1}^{(L)} V_S \right)^{2g-2}
\ee
which has left-moving $U(1)_{K3}$ charge $4g-4$. It must be accompanied by 
\be
\left(e^{\phi} G^{-}\right)^{4g-4} \non
\ee
which leaves $2g-2$ units of ghost charge. 

This leads to our selection rule for computing the $\tF^{(g)}(q)$ coupling.  The amplitude must be calculated on a surface with $\chi=2-2g$. In general,
this will involve contributions from both oriented as well as unoriented world-sheets with and without boundaries. 

It is important to note that our selection rule constraint applies to $\tF^{(g)}(q)$ with $g\geq 1$. The cases $g=0$ and to some extent $g=1$ are distinguished because those Riemann surfaces are not generic. We certainly expect at least a tree-level and $1$-loop contribution to the hypermultiplet metric determined by  $\tF^{(0)}(q)$. It is possible that there are no further perturbative contributions beyond $1$-loop (with an appropriate choice of fields) but this has yet to be established. 

We can also consider type I on $K3$ rather than $K3\times T^2$. In this case, the string coupling sits cleanly in a hypermultiplet. Under S-duality, this hypermultiplet maps to the $\vol(K3)$ of the dual heterotic string on $K3$ with the relation
\be
\left(g_s^I \right)^2 = \frac{1}{ V^H_{K3} } 
\ee
where $g_s^I$ is the $6$-dimensional type I coupling. This correlates the $\alpha'$ expansion in the heterotic string with the loop expansion in type I. Our selection rule then implies that the $\tF^{(g)}(q)$ with $g\geq 1$ computed in the heterotic string are generated at genus $g$ at a fixed order in $\alpha'$ perturbation theory together with world-sheet genus $g$ instantons. 

%%%%%%%%%%%%%%%%%%%%%%%%%%%%
%%%%%%%%%%%%%%%%%%%%%%%%%%%%

\section{Comments on the Topology of Instantons}
\label{topology}

We will now discuss some aspects of the instanton corrections to the $\tF^{(g)}(q)$ hypermultiplet couplings of~\eq{eq:hypercoup}. As we have described in some detail, sLag instantons in type IIA string theory on a $3$-fold $X$ which admits a $K3$ fibration should be characterized by a kind of genus. More precisely, we would like to conjecture a lower bound on the topological complexity of a sLag given in terms of the genus of the dual heterotic world-sheet instanton. 

Let us recall that  a Euclidean $D2$-brane  wrapping a sLag $L$ in $X$ will lead to couplings in the four-dimensional effective
action of the form 
\begin{equation}
\label{eq:instact}
 \delta S_{\text{eff}} \sim \int d^4 x~ N(q^i) \exp\left( -\frac{1}{g_s} \left| \int_L \Omega\right| + i \int_L C_3 \right) \left[\ldots\right],
 \end{equation}
where $\left[\ldots\right]$ denotes a particular space-time coupling of the low energy fields. Here $C_3$ is the pull-back of the Ramond-Ramond $3$-form to $L$. The factor $N(q^i)$ denotes the one-loop determinants from fluctuations about the semi-classical instanton configuration. If there is a moduli space of sLags, $N(q^i)$ will also include a (supersymmetric) integral over this moduli space. As described in section~\ref{reviewslag}, the moduli space of geometric deformations has real dimension $b_1(L)$. In type IIA, the moduli space is further enlarged by a choice of flat bundle on $L$. The corresponding moduli space has complex dimension $b_1(L)$. A reasonable measure of the topological complexity of a sLag is therefore provided by $b_1(L)$.

How is this measure related to the dual heterotic string instanton? The simplest approach is to correlate the moduli spaces of both kinds of instanton. For example, consider the duality chain described in figure~\ref{fig:dualitymap}. The starting point is a heterotic world-sheet instanton wrapping a curve $C$ embedded in a $K3$ surface $S$. To preserve $1/2$ of the supersymmetries, $C$ must be holomorphic in some complex structure. Based on the
argument in section \ref{sec:sLagK3}, this means that given any cycle $C$ with self-intersection $2g-2$ there will be a holomorphic world-sheet 
instanton of genus $g$ with class either $C$ or $-C$.  For $g> 0$, there is a moduli space of instantons. The virtual dimension of the moduli space for such an instanton is $2g$. These $2g$ bosonic zero modes correspond to holomorphic sections of the normal bundle $N$ of $C$ in $S$. 

Under the duality chain, we generally arrive at a bound state of NS5-branes and sLag Euclidean $D2$-brane instantons. For simplicity, let us assume that the NS5-brane charge is zero. Now there is no reason that the moduli space of the resulting sLag need agree with the moduli space of the heterotic instanton. The duality chain involves a strong coupling lift in the final step. Under S-duality, bundle and geometric moduli are typically reorganized. However on tracing through the duality sequence, we are led to conjecture that the resulting sLag moduli space must increase at least linearly with $g$. We will explain the origin of this conjecture from the counting fermion zero modes around each instanton configuration momentarily.  So we expect a relation of the form
\be
b_1(L) \geq a g + b
\ee
for some constants $a,b$. Using the results of~\cite{gross-1996}, we see that a $g=0$ curve maps to an $S^3$ sLag. This fixes $b=0$. Determining $a$ is more difficult but we will argue that $a=1$ so a lower bound on the topological complexity of the sLag resulting from a genus $g$ instanton is given by
\be \label{toprelation}
b_1(L) \geq  g . 
\ee        
It might be possible to sharpen this bound. For example, the analysis of~\cite{gross-1996}\ also shows that a particular elliptic curve in $S$ with $g=1$ maps to a $T^3$ sLag with $b_1 =3$.

\subsection{Heterotic world-sheet instantons}
We begin in the heterotic string where we want to describe the fermion zero modes for a genus $g$ world-sheet instanton which contributes to~\eq{eq:hypercoup}. Such a BPS instanton automatically has $6$ bosonic zero modes corresponding to the position of the instanton in $\RR^6$.  Accompanying these bosonic zero modes are $4$ right-moving fermion zero modes needed to construct the chiral superspace measure
\be
\int d^6x d^2\tha d^2 \overline{\ttha}. 
\ee
The superspace integral has already been implicitly performed in~\eq{eq:hypercoup}. 

Because the configuration is BPS, the collective coordinate dynamics is supersymmetric with $4$ supercharges.  This makes the counting 
of right-moving fermionic zero modes possible since they necessarily pair with the bosonic moduli: there are $4g$ additional right-moving fermion zero modes.

In addition, there are left-moving fermions that
couple to the restriction of the holomorphic vector bundle $E$, discussed in section~\ref{heteroticreview}, to $C$.  Since our vertex operators do not carry any gauge indices, the path-integral
over the left-moving sector contributes a factor to the correlator proportional to the Pfaffian of the (left-moving) 
Dirac operator twisted by $E|_C$.  In general, this factor is moduli-dependent and must be computed to obtain 
the exact form of the correction to the coupling. Let us set aside this non-universal left-moving factor for a moment to focus on the right-moving fermion zero modes. The structure of these zero modes will 
constrain which world-sheet instantons correct the hypermultiplet F-term couplings. 

To determine whether a given instanton can generate an $\tF^{(g)}$ coupling, we need to examine the coupling of the string world-sheet to the background supergravity fields. This will determine the coupling of the instanton to the scalars $(\phi_Z, \phi_S)$ of the universal hypermultiplet.  In physical gauge,\footnote{Physical gauge is a particular gauge-fixed form of the Green-Schwarz string.} the couplings of the string world-sheet fields are known to second order in fermions~\cite{Grisaru:1988jt, Grisaru:1988sa}. We note that the universal hypermultiplet contains $3$ scalars obtained by reducing the NS $B$-field on the anti-self-dual forms of $K3$. In ten dimensions, the relevant coupling of the world-sheet fermions to $H=dB$ is given by 
\be \label{backcoupling}
\int d^2 \sigma \overline{\Theta} \gam^{\mu ij}H_{\mu i j} \Theta. 
\ee
We have chosen an index structure such that $\mu$ is an index in $\RR^6$ and $i,j$ are holomorphic indices in $K3$. The $\Theta$ fields are world-sheet fermions valued in the target space spin-bundle. This provides a coupling of the world-sheet fermions to derivatives of the scalars $(\phi_Z, \phi_S)$ of the universal hypermultiplet.

To generate a contribution to the coupling~~\eq{eq:hypercoup}, the expansion of $e^{-S_{\rm inst}}$ in the background fields must contain a term of schematic form 
\be 
(\partial \phi_Z)^{2g-2} (\p \p \phi_S)^2 \non. \ee 
We see that bringing down powers of~\C{backcoupling}\ generates powers of $(\partial \phi_Z)$ coupled to two world-volume fermions. The $(\p \p \phi_S)^2$ is not visible at quadratic order in the fermions. There is, however, a direct analogy with the higher derivative vector couplings where a similar term is expected to appear at fourth order in fermions~\cite{Beasley:2005iu}. This term should have the form $(\bar{\Theta}\Theta)^4 (\p \p \phi_S)^2$, and we will assume it exists.

So in total, we need at least $2(2g-2) +4= 4g$ fermion zero modes in the instanton background to generate this space-time coupling. This is precisely the number of right-moving zero modes available in the background of a genus $g$ instanton. This is in agreement with the reasoning presented in section~\ref{ftermsheterotic} which, in addition, implies no higher genus instanton contributions. 

Even if the right-moving zero modes of the instanton allow a contribution to some $\tF^{(g)}$, that contribution might still vanish. For example, this will happen whenever the Pfaffian of the left-moving Dirac operator vanishes. In the 
genus $0$ case this reduces to the statement the $V$ must restrict to a trivial bundle on $C$~\cite{Distler:1987ee}. The corresponding condition for the genus $g>0$ case has yet to be cast in such simple terms because the bundle $E$ restricted to $C$ generally does not split into a sum of line bundles.

A simple case where we can study the vanishing of this Pfaffian for $g>0$ is the standard embedding, $E\simeq TS$. As we will show, the Pfaffian will be zero for all instantons in this case. With the standard embedding,  $E|_C$ splits into a sum of line bundles, $TC\oplus N$, and by adjunction
\begin{equation} 
E|_C \simeq TC\oplus T^{\ast}C \simeq \overline{K} \oplus K,
\end{equation}
where $K$ is the canonical bundle on $C$. The left-moving spinors are sections of
\begin{equation}
K^{1/2} \otimes E|_C \simeq \overline{K}^{1/2} \oplus \left( K^{1/2} \otimes K \right).
\end{equation}
From the Riemann-Roch theorem, the numbers of holomorphic sections of these bundles satisfy
\begin{eqnarray}
h^0(\overline{K}^{1/2}) & \ge & 2-2g, \nonumber\\
h^0(K^{1/2}\otimes K) & \ge & 2g -2.
\end{eqnarray}
Hence, there will always be additional zero modes from the left-moving fermions, and the world-sheet instantons will
not contribute to correlators without additional insertions of the left-movers.  In other words, there are no world-sheet instanton
corrections to the $\tF^{(g)}$ in the case of the standard embedding. This extends the known non-renormalization of the moduli space metric to these higher derivative couplings.

\subsection{$D2$-brane instantons}

We can now turn to the structure of fermion zero modes around a Euclidean $D2$-brane instanton wrapped on a sLag $L$. 
This will provide evidence for the relation~\C{toprelation}\ by showing that instantons satisfying the bound have a chance of contributing to the couplings~\eq{eq:hypercoup}. For simplicity, we will restrict to singly wrapped branes. A BPS $D2$-brane instanton automatically has $4$ bosonic zero modes corresponding to the position of the instanton in $\RR^4$.  Accompanying these bosonic zero modes are $4$ fermion zero modes needed to construct the chiral superspace measure. 

A sLag with $b_1>0$ will have additional fermion zero modes partnered with the bosonic moduli. Once again, the collective coordinate dynamics is supersymmetric with $4$ supercharges.  Therefore there are $4 b_1$ additional real fermion zero modes. As in the case of the heterotic string, these zero modes are needed to construct the couplings~\C{eq:hypercoup}. 

The world-volume action of a $D2$-brane coupled to $10$-dimensional supergravity background fields has been worked out to quadratic order in the world-volume fermions; for example, see~\cite{deWit:1998tk, Harvey:1999as}. The action takes the form, 
\begin{equation}
 S_{\text{D2}} = S_{\text{DBI}} + \int_{L} d^3\sigma \sqrt{h}  \bar{\Theta}(\sigma) \Gamma^{IJKL} F_{4 IJKL} (\Phi(\sigma)) \Theta(\sigma)
                  +\ldots,
\end{equation}
where $\Phi: L \to M^{10}  $ describes the embedding of $L$ in the target space $M^{10}$ while $F_4= dC_3$. The field $\Theta$ is a world-volume fermion while $h$ is the induced metric on $L$. It is clear that when this coupling is reduced on 
$M^{10} \simeq \RR^4 \times X$, we will generate world-volume interactions that will include  couplings to the scalars of the universal hypermultiplet
\be
\bar{\Theta}\gamma^\mu \Gamma^{ijk}  \Theta \, \p_\mu \phi_Z \Omega_{ijk}.
\ee
Once again, each factor of $\p \phi_Z$ appears with two fermions.   In addition, we again expect the $D2$-brane action to include terms of 
the schematic form $(\bar{\Theta}\Theta)^4 (\p\p \phi_S)^2$.

For a $D2$-brane wrapping $L$ to contribute to~\C{eq:hypercoup}, the instanton must therefore have at least $4g$ real fermion zero modes. Additional fermion zero modes do not necessarily kill the contribution of a given instanton. These extra modes may be absorbed by bringing down other interactions from the $D2$-brane action like the curvature of the instanton moduli space metric.  

Now we can connect this discussion with the heterotic fermion zero mode count. Via the duality map, we expect each heterotic world-sheet instanton that wraps a singular fiber of $S$ to map to a sLag $L$ in $X$ as discussed in section~\ref{SubHetIIA}. If such an instanton contributes to an $\tF^{(g)}$ coupling then we expect the corresponding $D2$-brane wrapping $L$ to contribute to this same space-time coupling. For this to be possible, the fermion zero mode analysis requires that 
\be\label{repeat}
b_1(L) \geq g. 
\ee
It should be possible to check this relation at the orientifold locus by generalizing the analysis of~\cite{gross-1996}. Finally, we should mention that whether type IIA on $X$ admits a type I dual or a more general $E_8 \times E_8$ heterotic dual, we still expect to find a dual sLag $L$ satisfying~\C{repeat}\ for each world-sheet instanton.

%%%%%%%%%%%%%%%%%%%%%%%%%%%%%%%%%%%%%%%%%%
%%%%%%%%%%%%%%%%%%
\section*{Acknowledgements}
It is a pleasure to thank P. Aspinwall, C. Beasley, D. Fox, J. Harvey, S. Katz ,  O. Lunin, E. Martinec, D.~R. Morrison, N. Nekrasov, R. Plesser,  C. Romelsberger, S. Vandoren and E. Zaslow for helpful discussions. We would also like to thank the referee for correcting an error in an earlier 
version of the manuscript. The work of N.~H.,  I.~M. and  S.~S. is supported in part by NSF CAREER Grant No. PHY-0094328 and by NSF Grant No. PHY-0401814. N.~H. is also supported in part by a Fermi-McCormick Fellowship. S.~S. and N.~H. would like to thank the Aspen Center for Physics for hospitality during the completion of this project. 

%%%%%%%%%%%%%%%%%%%%%%%%%%%%%%%%%%%%
%%%%%%%%%%%%%%%%%%%%%%%%%%%%%%%%%%%%
%\bibliographystyle{/Users/Nick/utphys} %\bibliography{/Users/Nick/myrefs}
%\bibliographystyle{utphys} \bibliography{hyperrefs-final}
%%%%%%%%%%%%%%%%%%%%%%%%%%%%%%%%%%%%

\providecommand{\href}[2]{#2}\begingroup\raggedright\endgroup

\end{document}